\documentclass[journal, hidelinks, 10pt]{IEEEtran}
\usepackage{amsmath,amsfonts,amssymb,mathtools}
\usepackage{csquotes}
\ifCLASSOPTIONcompsoc
	\usepackage[caption=false,font=normalsize,labelfont=sf,textfont=sf]{subfig}
\else
	\usepackage[caption=false,font=footnotesize]{subfig}
\fi
\usepackage{textcomp}
\usepackage{stfloats}
\usepackage{url}
\usepackage{verbatim}
\usepackage{algorithm}
\usepackage{siunitx}
\usepackage{graphicx}
\usepackage{cite}
\usepackage{orcidlink}
\usepackage{xfrac}
\usepackage[acronym]{glossaries}
\glsdisablehyper
\newacronym{wss}{WSS}{wide-sense stationary}
\newacronym{cs}{CS}{cyclostationary}
\newacronym{kl}{KL}{Karhunen-Loève}
\newacronym{cl}{CL}{Cramér-Loève}
\newacronym{psd}{PSD}{power spectral density}
\newacronym{caf}{CAF}{cyclic autocorrelation function}
\newacronym{sc}{SC}{spectral correlation}
\newacronym{dtft}{DTFT}{discrete-time Fourier transform}
\newacronym{lti}{LTI}{linear time-invariant}
\newacronym{snr}{SNR}{signal-to-noise ratio}
\newacronym{mse}{MSE}{mean squared error}
\newacronym{mmse}{MMSE}{minimum mean squared error}
\newacronym{fresh}{FRESH}{frequency-shift}
\newacronym{cwf}{CWF}{cyclic Wiener filter}
\newacronym{pam}{PAM}{pulse-amplitude modulation}
\usepackage{balance}
\usepackage{microtype}
\usepackage{scalerel}
\usepackage[english]{babel}
\usepackage[x11names]{xcolor}
\usepackage{tikz}
	\usetikzlibrary{arrows.meta}
	\usetikzlibrary{calc}
	\usetikzlibrary{decorations.text,calligraphy}	
	\usetikzlibrary{decorations.pathmorphing}
	\usetikzlibrary{decorations.pathreplacing}
	\usetikzlibrary{decorations.shapes}
	\usetikzlibrary{fadings}
	\usetikzlibrary{fit}
	\usetikzlibrary{patterns}
	\usetikzlibrary{quotes}
	\usetikzlibrary{shapes.arrows}
	\usetikzlibrary{shapes.geometric}
	\usetikzlibrary{shapes.misc}
	\usetikzlibrary{shapes.multipart}
	\usetikzlibrary{shapes.symbols}
	\usetikzlibrary{through}
	\usetikzlibrary{3d}
	\usetikzlibrary{external}
\usepackage{pgfplots}
	\pgfplotsset{compat=1.18}
	\usepgfplotslibrary{decorations.softclip}
	\usepgfplotslibrary{fillbetween}
	\usepgfplotslibrary{groupplots}
	\usepgfplotslibrary{units}
	\usepgfplotslibrary{colormaps}
	\usepackage{pgfplotstable}
	\usepackage{pgffor}
	\usepackage{pgfmath}
	\usepackage{listofitems}
	
\usepackage{fancyhdr}

\let\originalleft\left
\let\originalright\right
\renewcommand{\left}{\mathopen{}\mathclose\bgroup\originalleft}
\renewcommand{\right}{\aftergroup\egroup\originalright}

\newcommand{\corr}[1]{\ensuremath{\mathrm{R}_{#1}}}
\newcommand{\herm}[1]{\ensuremath{#1^{\mathrm{H}}}}
\newcommand{\trans}[1]{\ensuremath{#1^{\mathrm{T}}}}
\newcommand{\id}[1]{\ensuremath{\mathbf{I}_{#1}}}
\newcommand{\cycmat}[1]{\ensuremath{\mathbf{S}_{#1}}}
\newcommand{\klcycmat}[1]{\ensuremath{\boldsymbol{\mathcal{S}}_{#1}}}
\newcommand{\power}[1]{\ensuremath{\mathrm{P}_{#1}}}
\newcommand{\euler}{\ensuremath{\mathrm{e}}}
\newcommand{\imunit}{\ensuremath{\mathrm{j}}}
\newcommand{\corrmat}{\ensuremath{\mathbf{R}}}
\newcommand{\cyc}{\ensuremath{\mathrm{S}}}
\newcommand{\klcyc}{\ensuremath{\mathcal{S}}}
\newcommand{\snr}{\ensuremath{\mathrm{SNR}}}
\newcommand{\reals}{\ensuremath{\mathbb{R}}}
\newcommand{\complex}{\ensuremath{\mathbb{C}}}
\newcommand{\integs}{\ensuremath{\mathbb{Z}}}
\newcommand{\mmse}{\ensuremath{\mathrm{MMSE}}}
\newcommand{\coherence}{\ensuremath{\mathbf{C}}}
\newcommand{\der}{\ensuremath{\mathrm{d}}}

\DeclareMathOperator{\expec}{\mathrm{E}}
\DeclareMathOperator*{\argmax}{arg\,max}
\DeclareMathOperator{\Diag}{\mathrm{Diag}}
\DeclareMathOperator{\diag}{\mathrm{diag}}
\DeclareMathOperator{\kl}{\mathrm{KL}}
\DeclareMathOperator{\trace}{\mathrm{Tr}}
\DeclareMathOperator{\mse}{\mathrm{MSE}}

\DeclareFontFamily{U}{wncy}{}
\DeclareFontShape{U}{wncy}{m}{n}{<->wncyr10}{}
\DeclareSymbolFont{mcy}{U}{wncy}{m}{n}
\DeclareMathSymbol{\Sha}{\mathord}{mcy}{"58}

\newcommand{\ie}{\textit{i.e.} }
\newcommand{\eg}{\textit{e.g.} }
\makeatletter
    \newcommand{\pushright}[1]{\ifmeasuring@#1\else\omit\hfill$\displaystyle#1$\fi\ignorespaces}
\makeatother

\newtheorem{theorem}{Theorem}
\newtheorem{corollary}{Corollary}
\newtheorem{proposition}{Proposition}

\newcommand{\changefont}{
	\color{blue}\fontsize{9}{9}\selectfont
}

\let\oldmaketitle\maketitle
\renewcommand{\maketitle}{%
	\oldmaketitle
	\thispagestyle{fancy}
}

\begin{document}

	\title{Asymptotic Analysis of Synchronous Signal Processing}
	
	\author{
		Marc~Vilà-Insa\,\orcidlink{0000-0002-7032-1411},
		and Jaume~Riba\,\orcidlink{0000-0002-5515-8169}, \IEEEmembership{Senior Member, IEEE}%
		\thanks{This work was (partially) funded by project RODIN (PID2019-105717RB-C22) by MICIU/AEI/10.13039/501100011033, project MAYTE (PID2022-136512OB-C21) by MICIU/AEI/10.13039/501100011033 and ERDF/EU, grant 2021 SGR 01033 and grant 2022 FI SDUR 00164 by Departament de Recerca i Universitats de la Generalitat de Catalunya.}%
		\thanks{The authors are with the Signal Processing and Communications Group (SPCOM), Departament de Teoria del Senyal i Comunicacions, Universitat Politècnica de Catalunya (UPC), 08034 Barcelona, Spain (e-mail: \{marc.vila.insa, jaume.riba\}@upc.edu).}
	}
	
	\maketitle
	
	\begin{abstract}
		This paper extends various theoretical results from stationary data processing to \acrfull{cs} processes under a unified framework.
		We first derive their asymptotic eigenbasis, which provides a link between their Fourier and \acrfull{kl} expansions, through a unitary transformation dictated by the cyclic spectrum.
		By exploiting this connection and the optimalities offered by the \acrshort{kl} representation, we study the asymptotic performance of smoothing, filtering and prediction of \acrshort{cs} processes, without the need for deriving explicit implementations.
		We obtain \acrlong{mmse} expressions that depend on the cyclic spectrum and include classical limits based on the power spectral density as particular cases.
		We conclude this work by applying the results to a practical scenario, in order to quantify the achievable gains of synchronous signal processing.
	\end{abstract}
	
	
	\section{Introduction}\label{sec:intro}
		\IEEEPARstart{M}{any} random phenomena in nature and engineering exhibit statistical properties that vary periodically with time.
Notable examples are encountered in Earth sciences (\eg hydrology, oceanography and climatology~\cite{Napolitano2016}) as well as in human activities (\eg econometrics, electronic design and mechanical engineering~\cite{Napolitano2016}).
Of particular interest are the signals used in digital communications and radar systems, where periodicities occur due to the use of constant symbol rates and carrier frequencies~\cite[Ch.~12]{Gardner1988}.

These cyclic behaviors are typically modeled statistically as \textit{\acrfull{cs}} or \textit{periodically correlated} stochastic processes~\cite{Madisetti2010}.
The research on signal processing methods that exploit their unique characteristics is broad, rich and mature~\cite{Gardner2006}.
Numerous applications have emerged in areas in which these phenomena typically arise, which include signal extraction/separation from spatial or temporal mixtures~\cite[Ch.~14]{Gardner1988}, modeling~\cite[Ch.~7]{Napolitano2012}, blind system equalization and identification~\cite{YesteOjeda2019}, spectrum sensing~\cite{Chopra2016} and time-delay estimation~\cite{Madisetti2010}.
They are usually based on exploiting the diversity offered by such structured statistical change, and their performance is generally affected by the accuracy in prior knowledge of the inherent periods of variation.

Concerning linear filtering, it is well known that the \textit{\acrfull{cwf}}~\cite{Gardner1993} is the optimal periodically time-variant linear estimator for \acrshort{cs} signals.
It can be understood as a combination of several linear shift-invariant filters operating on replicas of the signal of interest under different frequency shifts (\acrshort{fresh})~\cite[Ch.~10]{Schreier2010}, thus providing a variety of improved designs~\cite{Chopra2016}.
Compared to simpler methods that assume stationarity, detectors and estimators derived from \acrshort{cs} modeling perform better at discriminating signal from interference, by being aware of the intrinsic period of the signals involved, either the one of interest~\cite{Riba2014, Ramirez2015}, or the one considered interference~\cite{Chien2020,Elgenedy2018}.

The derivation of asymptotic performance limits of \acrfull{lti} filters with unconstrained length has been a fundamental tool within the classical theory of stationary signals.
For a given problem, they provide the ultimate achievable performance in advance and inform about how far a constrained implementation is from the optimum, thus shedding light on the achieved complexity/performance trade-off.
Not less important is their role in unveiling key physical aspects of the signal that are the most relevant for the pursued processing performance.

In this sense, several bounds are well-known for stationary processes, on account of their asymptotic uncorrelatedness through Fourier analysis.
This property, which is in fact a corollary of the more general \textit{Szegö limit theorems}, comes from the asymptotic diagonalization of Toeplitz matrices through the unitary Fourier basis~\cite{Gray2005}.
Examples of these limits are expressions of \acrfull{mmse} through integral operators involving spectral coherences (\ie Wiener smoothing in the frequency domain~\cite[Sec.~12.3]{Oppenheim2017}).
The \textit{Kolmogorov-Szegö formula}~\cite[Ch.~6]{Vaidyanathan2007}
on the \acrshort{mmse} for asymptotic one-step linear prediction is another remarkable limit expression, and involves integral operators over the log-spectrum.
When causality is a design restriction, such as in applications requiring small processing latency, causal Wiener filtering becomes relevant.
The performance limit in this case involves, as in linear prediction, integral operators on the log-spectrum~\cite{Anderson1971}.
Since these causal formulas resemble Shannon capacity expressions, information theoretic interpretations have been explored in the literature~\cite{Kadota1971}.
A fundamental result, known as the \textit{I-\acrshort{mmse} formula} (or \textit{Guo-Shamai-Verdú (GSV) theorem}), was obtained in~\cite{Guo2005} and establishes a direct path between the \acrshort{mmse} of causal and non-causal processing for a very general kind of filtering.

Given the prevalence of \acrshort{cs} signals in both theoretical~\cite{Kipnis2018} and applied~\cite{Pries2018} research fields, one would expect to find results parallel to the ones mentioned for stationary data.
However, this is mostly not the case, possibly due to the lack of more general asymptotic properties of their autocorrelation structures.
Therefore, the main purpose of this paper is going in the direction of filling this gap and providing new insights within the context of filtering \acrshort{cs} signals.
The goal is to understand the ultimate performance gain achievable from synchronous signal processing, under the assumption of attainable synchronization with the periodic phenomenon that is inherent to the desired noisy data.

Classical treatments of \acrshort{cs} signals are usually based on their \textit{\acrfull{cl}} spectral expansion~\cite[Ch.~10]{Schreier2010}, which provides a direct physical interpretation of the problem at hand.
In many cases, this involves expressing them in terms of stationary components~\cite[Sec.~3.10]{Gardner2006}.
A prominent example is the \textit{harmonic series representation}~\cite{Gardner1975} or \textit{Gladyshev decomposition}~\cite[Sec.~7.2]{Hurd2007}, which represents a $P$-periodic \acrshort{cs} process as a sum of $P$ stationary signals with non-overlapping bands~\cite[Sec.~3.10]{Gardner2006}.
A different strategy entails converting a scalar \acrshort{cs} process into multivariate low-rate stationary data through the \textit{time series representation}~\cite[Sec.~12.6]{Gardner1986} or \textit{polyphase decomposition}~\cite[Sec.~17.2]{Madisetti2010}.
Although these techniques are useful for implementation purposes, they often blur the connection between performance bounds and physical parameters, and lead to processes with correlated spectral increments (when applied to non-stationary data).
In contrast, we follow a fully uncorrelated approach based on the asymptotic eigendecomposition of periodic Toeplitz matrices~\cite{Riba2022}.
We leverage well-known results for stationary data and extend them without breaking the inherent structure of the signals, such that the resulting formulas remain interpretable.

Moreover, we capitalize on the general GSV theorem to explore the causal \acrshort{cwf}, instead of studying the causal constraint specific to the \acrshort{cs} problem.
By following this alternative route, we obtain mathematical expressions of performance bounds that depend on the cyclic spectrum explicitly, thus becoming insightful generalizations that contain classical limits based on the spectrum as particular cases.
The tools developed in the paper allow to obtain theoretical results for any particular application: they provide knowledge from the problem without the need for deriving explicit implementations.
The exposition is probabilistic and focused on discrete-time signals and systems encountered in signal processing and communications, while maintaining a continuous spectrum interpretation.
As a result, it is possible to plot performance limits against the relevant signal parameters in a very direct manner, as integral expressions. 

The main contributions of this paper are listed below:
\begin{itemize}
	\item We revisit and refine results obtained in~\cite{Riba2022}.
	Using an incremental formulation~\cite[Ch.~8]{Schreier2010}, we obtain the asymptotic \textit{\acrfull{kl}} integral expansion of \acrshort{cs} processes.
	This allows it to be interpreted as a decorrelation of their \acrshort{cl} representation, bridging them through a unitary transformation.
	\item We extend classical results of stationary filtering to \acrshort{cs} processes.
	In particular, we provide a new sense of optimality for the \acrshort{cwf} by proving its equivalence to the Wiener filter in the \acrshort{kl} domain, which exploits the energy compaction properties of such representation.
	\item We develop a unified theoretical analysis of smoothing, filtering and prediction of \acrshort{cs} processes thanks to the discrete-time approach and the use of \cite[Th.~8]{Guo2005} to connect the three regimes.
	We quantify the \acrshort{mmse} and maximum achievable gain with synchronous processing in each case.
	\item We relate the obtained \acrshort{mmse} expressions to the \textit{spectral coherence} and \textit{coherence matrix}~\cite{Ramirez2022} in an explicit manner.
\end{itemize}

The text is organized as follows.
Section~\ref{sec:prelim} establishes the basic notions and concepts used in the rest of the derivations.
Section~\ref{sec:kl} develops the theoretical results regarding the \acrshort{kl} expansion of \acrshort{cs} processes, as well as its connection to the \acrshort{cl} expansion and other properties of interest.
The filtering problem is explored throughout the rest of the paper: Section~\ref{sec:wiener} presents the model, while Sections~\ref{sec:noncausal} to~\ref{sec:prediction} study various asymptotic performance metrics for smoothing, filtering and prediction, respectively.
Finally, some numerical illustrations of the theoretical contents are displayed in Section~\ref{sec:numerical}.

\textit{Notation:} Vectors and matrices are denoted by boldface lowercase and uppercase letters.
An element $(r,c)$ from a matrix $\mathbf{A}$ is indicated by $[\mathbf{A}]_{r,c}$.
The conjugate, transpose and conjugate transpose operators are $\cdot^*$, $\trans{\cdot}$ and $\herm{\cdot}$, respectively.
The inverse of $\mathbf{A}$ is $\mathbf{A}^{-1}$.
The trace of $\mathbf{A}$ is $\trace[\mathbf{A}]$, while its determinant is $\lvert\mathbf{A}\rvert$.
The identity matrix of size $N$ is $\id{N}$, and an all-zeros vector of size $M$ is $\mathbf{0}_M$.
A product of matrices that depend on the same parameter $\mathbf{A}(s)\mathbf{B}(s)\cdots\mathbf{C}(s)$ is shortened by $(\mathbf{AB}\cdots\mathbf{C})(s)$.
Operator $\Diag(a,b,\dots,c)$ constructs a diagonal matrix from the set $\{a,b,\dots,c\}$, while $\diag(\mathbf{A})$ performs the opposite action.
The imaginary unit is $\imunit$.
The expectation operator is $\expec[\cdot]$.
Dirac delta is $\delta(s)$ and Kronecker delta is $\delta_{a}$.
The Dirac comb with separation $T$ is $\Sha_{T}(s)$.
If a set $a$ \textit{majorizes} another set $b$, it is indicated as $a\succ b$.
The element-wise product between matrices is $\circ$.

	\section{Preliminary definitions and background}\label{sec:prelim}
		\noindent Let $\{x(n)\in\complex\}$ be a discrete-time random process with time index $n\in\integs$.
For a clearer exposition, all the considered processes will be assumed zero-mean and \textit{proper} (\ie $\expec[x(n)x(n+m)]=0$, $\forall n,m\in\integs$)~\cite[Sec.~1.6.2]{Schreier2010}.
Although the study is general, this has been done to avoid the treatment of complementary statistical descriptions that are required for a complete characterization of improper signals~\cite[Sec.~1.7]{Schreier2010}, in favor of clarity of exposition and simplicity.

The autocorrelation function of $\{x(n)\}$ is defined as
\begin{equation}\label{eq:autocorr}
    \corr{x}(n,m)\triangleq\expec\bigl[x(n+m)x^{*}(n)\bigr],
\end{equation}
where index $m$ is referred to as \textit{lag}.
If $\expec[\lvert x(n)\rvert^2]<\infty$ for all $n$, $\{x(n)\}$ is known as \textit{second-order}~\cite[Ch.~1]{Hurd2007}.
Various classes of random processes can be identified based on properties of their autocorrelation.
In particular, \acrshort{cs} processes display a periodic structure with period $P$ within it:
\begin{equation}
	\corr{x}(n,m)\equiv\corr{x}(n+lP,m),\quad\forall l\in\integs.\label{eq:cs_def}
\end{equation}
When $P=1$, a \acrshort{cs} process becomes \textit{\acrfull{wss}}, whose autocorrelation function is independent from the time index: $\corr{x}(n,m)\equiv\corr{x}(m)$ (\ie it displays \textit{shift-invariant second-order statistics}~\cite[Ch.~8]{Schreier2010}).

The following definition will prove useful in the sequel.
The \textit{cyclic spectrum} is the two-dimensional Fourier transform of the autocorrelation function, both in the time and lag domains.
Particularized for \acrshort{cs} processes, it is expressed as
\begin{equation}
    \cyc_{X}^{(\frac{k}{P})}(f)=\frac{1}{P}\smashoperator[l]{\sum_{n=0\vphantom{\in\integs}}^{P-1}}\smashoperator[r]{\sum_{m\in\integs}}\corr{x}(n,m)\euler^{-\imunit2\pi(\frac{kn}{P}+fm)},\label{eq:cyc_spec}
\end{equation}
for all $k\in\{0,\dots,P-1\}$.
This is known as the \textit{cyclic Wiener-Khinchin relationship}~\cite[Ch.~1,~(39a)]{Gardner1994}.

While these expressions provide a statistical representation of the properties of $\{x(n)\}$, in many signal processing applications it is convenient to have a spectral representation of the process itself.
In the following sections, the \acrshort{cl} and \acrshort{kl} expansions are presented.

\subsection{CL spectral representation}\label{ssec:cl}
    
    A complex second-order random process $\{x(n)\}$ can be expressed as
    \begin{equation}\label{eq:cl}
        x(n) = \int_{\mathrlap{0}}^{\mathrlap{1}} \euler^{\imunit2\pi fn} \der\nu_x(f),
    \end{equation}
    where $\der\nu_x(f)$ are the increments of $\{\nu_x(f)\}_{f\in[0,1)}$, which is a random process in the spectral domain\footnote{This entity is usually interpreted as a \textit{random spectral measure} (\eg see~\cite[Sec.~5.2]{Hurd2007}), and is also known as \textit{integrated spectrum} of $\{x(n)\}$~\cite[Sec.~1.1.2]{Napolitano2012}.}.
    This Riemann-Stieltjes integral representation is known as the \acrshort{cl} expansion, and processes that can be expressed in such a way are called \textit{harmonizable}~\cite[Ch.~5]{Hurd2007}.
    
    The \acrshort{cl} expansion is of great conceptual interest since it decomposes $\{x(n)\}$ onto the Fourier basis (complex exponentials), thus preserving the frequency interpretation of the transformed domain.
    If the derivative process $X(f)\triangleq\frac{\der\nu_x(f)}{\der f}$ existed, it would be the \acrfull{dtft} of $\{x(n)\}$~\cite[Eq.~(4.13)]{Wang2012}.
    However, $\{x(n)\}$ is not square-summable in general and its \acrshort{dtft} might be ill-defined.
    Thus we resort to the increment definition~\cite[Prop.~5.12]{Hurd2007}:
    \begin{equation}\label{eq:icl}
        \der\nu_x(f) = \smashoperator[l]{\lim_{N\to\infty}} \frac{1}{N} \smashoperator{\sum_{n=-\frac{N}{2}}^{\frac{N}{2}-1}} x(n) \euler^{-\imunit2\pi fn},
    \end{equation}
    where we assume $N\in2\mathbb{N}$ without loss of generality.
    
    In a similar manner to~\eqref{eq:autocorr}, a correlation function between frequentially displaced spectral increments can be defined~\cite[Sec.~9.2]{Schreier2010}:
    \begin{equation}\label{eq:spec-corr}
        \cyc_{X}(\alpha,f) \der\alpha \der f \triangleq \expec\bigl[\der\nu_x(f)\der\nu^{*}_x(f-\alpha)\bigr],
    \end{equation}
    known as \textit{spectral correlation} or \textit{Loève spectrum}.
    When $\alpha=0$, it is convenient to define the \textit{\acrfull{psd}}~\cite[Sec.~9.2.2]{Schreier2010}:
    \begin{equation}\label{eq:psd}
        \cyc_{X}(f) \der f \triangleq \expec\bigl[\lvert\der\nu_x(f)\rvert^2\bigr].
    \end{equation}
    Note that $\cyc_{X}(\alpha,f)$ represents \textit{power over frequency squared} (\ie 2-dimensional power density), whereas $\cyc_{X}(f)$ represents \textit{power over frequency} (\ie 1-dimensional power density), implying that $\cyc_{X}(0,f)\der\alpha=\cyc_{X}(f)$.

    Many stochastic processes can be described from properties of their \acrshort{cl} spectral representation.
    For instance, \acrshort{wss} processes display the unique characteristic of having orthogonal increments~\cite[Sec.~8.1]{Schreier2010}, which translates into
    \begin{equation}\label{eq:orth}
        \cyc_{X}^{(\mathrm{WSS})}(\alpha,f) = \cyc_{X}(f)\delta(\alpha).
    \end{equation}
    On the contrary, \acrshort{cs} processes present the following spectral correlation~\cite[Sec.~10.1.1]{Schreier2010}:
    \begin{equation}\label{eq:cs_spec_corr}
        \cyc_{X}^{(\mathrm{CS})}(\alpha,f)  = \smashoperator{\sum_{k=0}^{P-1}} \cyc_{X}^{(\frac{k}{P})}(f) \delta\bigl(\alpha-\tfrac{k}{P}\bigr),
    \end{equation}
    in terms of the cyclic \acrshort{psd} from~\eqref{eq:cyc_spec}.
    
    Representing a non-\acrshort{wss} process over the Fourier domain comes at the cost of losing the orthogonality property~\eqref{eq:orth}.
    For this reason, it might be useful to study an alternative spectral expansion that preserves it.

	\section{KL representation of CS processes}\label{sec:kl}
		\noindent This section is concerned with a spectral representation of a second-order random process known as \acrshort{kl} expansion.
It is commonly used to express a continuous-time stochastic process as an infinite series of discrete spectral random variables (\eg see Eq.~(9.6) in~\cite[Sec.~9.1]{Schreier2010}).
Nevertheless, the variant relating a continuous-spectrum process to a discrete-time one, more akin to integral forms in~\cite[Ch.~9]{Doob1990} and~\cite[Sec~1.1.5]{Napolitano2012}, is the one that will be discussed throughout this work.
Its use and interpretation are completely analogous to the former one.

Given a second-order random process $\{x(n)\}$, its \acrshort{kl} expansion is defined as the following \textit{Doob integral}~\cite[Sec.~3.2.3]{Bremaud2014}:
\begin{equation}\label{eq:ikl}
	x(n)= \int_{\mathrlap{0}}^{\mathrlap{1}} \phi(n,\lambda) \der\xi_x(\lambda),
\end{equation}
such that $\{\der\xi_x(\lambda)\}_{\lambda\in[0,1)}$ are increments of a random process in the \acrshort{kl} spectral domain.
The set $\{\phi(n,\lambda)\}$, for all $n\in\integs$ and $\lambda\in[0,1)$, contains eigenfunctions that satisfy the \textit{orthonormality condition}~\cite[Def.~2.17]{Kennedy2013} and \textit{completeness relation}~\cite[Th.~2.13]{Kennedy2013}, which are
\begin{equation}\label{eq:unitary}
	\begin{cases}
		\int_{\mathrlap{0}}^{\mathrlap{1}}\phi(n,\lambda)\phi^*(n',\lambda)\der\lambda &=\delta_{n-n'}\\
		\sum_{n\in\integs}\phi(n,\lambda)\phi^*(n,\lambda') &=\delta(\lambda-\lambda')
	\end{cases},
\end{equation}
respectively.
They are obtained by solving the succeeding eigenequation\footnote{This eigenproblem is usually set as an integral equation, such as in~\cite[Eq.~(6.1)]{vanTrees2013} or in~\cite[Eq.~(9.5)]{Schreier2010}. A well-known case in summation form can be found in~\cite[Eq.~(18)]{Slepian1978}, whose solution is the set of \textit{Slepian functions} (\ie \textit{prolate spheroidal functions}).}:
\begin{equation}\label{eq:eigenequation}
	\sum_{l\in\integs}\corr{x}(l,k-l)\phi(l,\lambda)=\klcyc_{\mathcal{X}}(\lambda)\phi(k,\lambda).
\end{equation}
The term $\klcyc_{\mathcal{X}}(\lambda)$ is the \textit{\acrshort{kl} spectrum} of $\{x(n)\}$.
By combining~\eqref{eq:unitary} and~\eqref{eq:eigenequation}, it is straightforward to obtain explicit conversion formulas:
\begin{equation}\label{eq:eigendecomposition}
	\begin{aligned}
		\corr{x}(l,k-l) &= \int_{\mathrlap{0}}^{\mathrlap{1}} \phi(k,\lambda) \klcyc_{\mathcal{X}}(\lambda) \phi^*(l,\lambda) \der\lambda\\
		\klcyc_{\mathcal{X}}(\lambda) \delta(\lambda-\lambda') &= \smashoperator{\sum_{k,l\in\integs}} \phi(l,\lambda)\corr{x}(l,k-l)\phi^*(k,\lambda').
	\end{aligned}
\end{equation}
Complementary to~\eqref{eq:ikl}, the \acrshort{kl} transform\footnote{Expression~\eqref{eq:ikl} is sometimes referred to as \textit{backward \acrshort{kl} transform} or \textit{synthesis}. Similarly, \textit{forward \acrshort{kl} transform} or \textit{analysis} are used for~\eqref{eq:kl}.} of $\{x(n)\}$ is obtained by projecting the random process onto its \acrshort{kl} basis $\{\phi(n,\lambda)\}$:
\begin{equation}\label{eq:kl}
	\der\xi_x(\lambda) = \smashoperator[l]{\lim_{N\to\infty}} \frac{1}{N} \smashoperator{\sum_{n=-\frac{N}{2}}^{\frac{N}{2}-1}} x(n) \phi^*(n,\lambda) \triangleq \kl\bigl\{x(n)\bigr\}(\lambda).
\end{equation}
The first equality can be checked by adapting the proof to~\cite[Prop.~5.12]{Hurd2007} to the set of \acrshort{kl} eigenfunctions and assuming $\sfrac{1}{N}\to\der\lambda$ as $N\to\infty$.

The defining characteristic of the \acrshort{kl} expansion is the orthogonality between spectral increments, \ie
\begin{equation}
    \begin{aligned}
    	\klcyc_{\mathcal{X}}(\beta,\lambda)\der\beta\der\lambda&\triangleq\expec\bigl[\der\xi_x(\lambda)\der\xi_x^{*}(\lambda-\beta)\bigr]\\
    	&=\klcyc_{\mathcal{X}}(\lambda)\delta(\beta)\der\beta\der\lambda,
    \end{aligned}\label{eq:kl-psd}
\end{equation}
which can be derived from~\eqref{eq:kl} and~\eqref{eq:eigendecomposition}.
This property generalizes~\eqref{eq:orth} and will prove to be fundamental in subsequent derivations of simple \acrshort{mmse} expressions.
The trade-off for spectral uncorrelatedness is the fact that each class of random processes has a different eigenbasis, compared to the simplicity of the \acrshort{cl} representation.
\acrshort{wss} processes are special in this sense because their \acrshort{kl} and \acrshort{cl} decompositions coincide and are both asymptotically given by the orthonormal Fourier basis.

The \acrshort{kl} representation of \acrshort{cs} processes of period $P$ was obtained in~\cite{Riba2022}.
To define it, the full \acrshort{kl} domain $[0,1)$ is segmented into $P$ non-overlapping domains and $\lambda$ is replaced by the pair $(p,\sigma)$.
The continuous variable $\sigma$ is defined in a spectral sub-band $[0,\sfrac{1}{P})$ and $p\in\{0,1,\dots,P-1\}$ is an index that selects one of these subintervals, \ie $\lambda\triangleq\sigma+\sfrac{p}{P}$.
\begin{theorem}{(Riba \& Vilà-Insa~\cite[Th.~1]{Riba2022})}\label{thm:cs}
    The \acrshort{kl} eigenbasis of a \acrshort{cs} process of period $P$ takes the form:
    \begin{equation}\label{eq:cs_basis}
		\phi_{\mathrm{CS}}^{(p)}(n,\sigma)=\sum_{q=0}^{P-1}b_{X}^{(p)}(q,\sigma)\euler^{\imunit2\pi(\sigma+\frac{q}{P})n}.
    \end{equation}
    Weighting coefficients $b_{X}^{(p)}(q,\sigma)$ are grouped in vectors
    \begin{equation}
	    \mathbf{b}_X^{(p)}(\sigma)\triangleq\trans{\bigl[b_{X}^{(p)}(0,\sigma),\dots,b_{X}^{(p)}(P-1,\sigma)\bigr]},
    \end{equation}
    and obtained by solving the following eigenproblem:
    \begin{equation}
        \cycmat{X}(\sigma)\mathbf{b}_{X}^{(p)}(\sigma)=\klcyc_{\mathcal{X}}^{(p)}(\sigma)\mathbf{b}_{X}^{(p)}(\sigma).\label{eq:eig_cyc_psd}
    \end{equation}
    Terms $\klcyc_{\mathcal{X}}^{(p)}(\sigma)\triangleq\klcyc_{\mathcal{X}}(\sigma+\sfrac{p}{P})$ are the eigenvalues of the Hermitian non-negative definite cyclic \acrshort{psd} matrix of the process, $\cycmat{X}(\sigma)\in\complex^{P\times P}$, given by
    \begin{equation}
        [\cycmat{X}(\sigma)]_{r,c}\triangleq \cyc_{X}^{(\frac{r-c}{P})}\bigl(\sigma+\tfrac{r}{P}\bigr).
    \end{equation}
\end{theorem}
\begin{IEEEproof}
    See Appendix~\ref{app:cs}.
\end{IEEEproof}

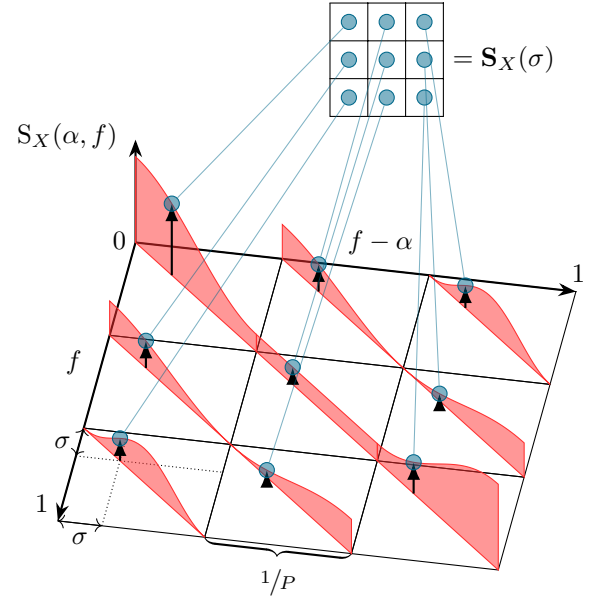
\begin{figure}[t]
	\centering
	\begin{tikzpicture}[dirac/.style={thick,arrows=-Latex[]}, cercle/.style={circle, minimum size=0.2cm, inner sep=0, fill=DeepSkyBlue4, fill opacity=0.5, draw=DeepSkyBlue4, solid}]
	\pgfmathsetmacro{\xmax}{3}
	\pgfmathsetmacro{\selx}{0.3}
	\begin{axis}[
		xmin=0,	xmax=1,
		ymin=0,	ymax=1,
		zmin=0,	zmax=3,
		view={100}{70},
		area plot/.style={
			fill opacity=0.5,
			draw=Firebrick1,
			fill=Firebrick1,
			mark=none,
			samples=20,
			samples y=1,
			smooth,
			domain=0:1/\xmax
		},
		xlabel={$f$},
		ylabel={$f-\alpha$},
		zlabel={$\cyc_{X}(\alpha,f)$},
		xtick={0,1},
		ytick={0,1},
		ztick={0},
		hide obscured z ticks=false,
		axis lines=center,
		major tick length=0,
		axis line style={-Stealth[], black, thick},
		every x tick label/.style={anchor=near xticklabel opposite},
		every axis x label/.style={
			at={(ticklabel cs:0.5)},
			anchor=near xticklabel opposite,
			xshift=-0.2em
		},
		every y tick label/.style={anchor=near yticklabel opposite},
		every axis y label/.style={
			at={(ticklabel cs:0.55)},
			anchor=near yticklabel opposite,
			yshift=0.3em
		},
		every z tick label/.style={yshift=0.5em},
		every axis z label/.style={
			at={(ticklabel cs:1)},
			anchor=south east,
			xshift=0.75em,
		},
		clip=false,
		use units=false
		]
		\addplot3[mesh, thin, black, domain=0:1, samples=\xmax+1, domain y=0:1, samples y=\xmax+1, forget plot] {0};
		\draw[thin, black, densely dotted] (1,\selx/\xmax) -- ++(-1/\xmax,0);
		\draw[thin, black, densely dotted] ({(\xmax-1+\selx)/\xmax},0) -- ++(0,1/\xmax);
		\draw[decorate, decoration={calligraphic brace, amplitude=4}, color=black, thick] (1,2/\xmax) -- ++(0,-1/\xmax) node[pos=0.5,below=5,black]{$\sfrac{1}{P}$};
		\draw[arrows=<->, thin, black, below] (1,0) -- ++(0,\selx/\xmax) node[pos=0.5] {$\sigma$};
		\draw[arrows=<->, thin, black, left] ({(\xmax-1)/\xmax},0) -- ++(\selx/\xmax,0) node[pos=0.5] {$\sigma$};
		\addplot3[area plot] (x,x,{2*cos(deg(pi*x*\xmax/2))^2 + 0.5}) \closedcycle;
		\draw[dirac] (\selx/\xmax,\selx/\xmax,0) -- ++(0,0,{2*cos(deg(pi*\selx/2))^2 + 0.5}) node (A1){};
		\addplot3[area plot] (x,{x+1/\xmax},{cos(deg(pi*x*\xmax/2))^2}) \closedcycle;
		\draw[dirac] (\selx/\xmax,{(\selx+1)/\xmax},0) -- ++(0,0,{cos(deg(pi*\selx/2))^2}) node (B1){};
		\addplot3[area plot]
		(x,{x+2/\xmax},{sin(deg(pi*x*\xmax))^2}) \closedcycle;
		\draw[dirac] (\selx/\xmax,{(\selx+2)/\xmax},0) -- ++(0,0,{sin(deg(pi*\selx))^2}) node (C1){};
		\addplot3[area plot]
		({x+1/\xmax},x,{cos(deg(pi*x*\xmax/2))^2}) \closedcycle;
		\draw[dirac] ({(\selx+1)/\xmax},\selx/\xmax,0) -- ++(0,0,{cos(deg(pi*\selx/2))^2}) node (A2){};
		\addplot3[area plot]
		({x+1/\xmax},{x+1/\xmax},0.5) \closedcycle;
		\draw[dirac] ({(\selx+1)/\xmax},{(\selx+1)/\xmax},0) -- ++(0,0,0.5) node (B2){};
		\addplot3[area plot]
		({x+1/\xmax},{x+2/\xmax},{sin(deg(pi*x*\xmax/2))^2}) \closedcycle;
		\draw[dirac] ({(\selx+1)/\xmax},{(\selx+2)/\xmax},0) -- ++(0,0,{sin(deg(pi*\selx/2))^2}) node (C2){};
		\addplot3[area plot]
		({x+2/\xmax},x,{sin(deg(pi*x*\xmax))^2}) \closedcycle;
		\draw[dirac] ({(\selx+2)/\xmax},\selx/\xmax,0) -- ++(0,0,{sin(deg(pi*\selx))^2}) node (A3){};
		\addplot3[area plot]
		({x+2/\xmax},{x+1/\xmax},{sin(deg(pi*x*\xmax/2))^2}) \closedcycle;
		\draw[dirac] ({(\selx+2)/\xmax},{(\selx+1)/\xmax},0) -- ++(0,0,{sin(deg(pi*\selx/2))^2}) node (B3){};
		\addplot3[area plot]
		({x+2/\xmax},{x+2/\xmax},{2*sin(deg(pi*x*\xmax/2))^2 + 0.5}) \closedcycle;
		\draw[dirac] ({(\selx+2)/\xmax},{(\selx+2)/\xmax},0) -- ++(0,0,{2*sin(deg(pi*\selx/2))^2 + 0.5}) node (C3){};
	\end{axis}
	\node (matrixorigin) at (3.6,6) {};
	\draw[step=0.5, line cap=rect, xshift=.1cm] (matrixorigin) grid ++(1.5,1.5);
	\draw[] ($(matrixorigin) + (0,1.5)$) -- ++(0,-1.5) -- ++(1.5,0);
	\node foreach \lloc in {A1, B1, C1, A2, B2, C2, A3, B3, C3}
	[cercle] (c\lloc) at (\lloc) {};
	\foreach \x/\y/\nom in {0.25/0.25/A3, 0.75/0.25/B3, 1.25/0.25/C3, 0.25/0.75/A2, 0.75/0.75/B2, 1.25/0.75/C2, 0.25/1.25/A1, 0.75/1.25/B1, 1.25/1.25/C1}
	\node[cercle] (d\nom) at ($(matrixorigin) + (\x,\y)$) {};
	\foreach \nom in {A1, B1, C1, A2, B2, C2, A3, B3, C3}
	\draw[thin,solid,DeepSkyBlue4,opacity=0.5] (c\nom) -- (d\nom);
	\node[right] at ($(matrixorigin) + (1.5,0.75)$) {$=\cycmat{X}(\sigma)$};
\end{tikzpicture}
	\caption{Graphical representation of the construction of $\cycmat{X}(\sigma)$ from the spectral correlation $\cyc_{X}(\alpha,f)$ of a \acrshort{cs} process, which is only defined across $\delta$-ridges spaced by $\sfrac{1}{P}$ both horizontally and vertically~\cite[Sec.~10.1.1]{Schreier2010}.}
	\label{fig:psd}
\end{figure}
The cyclic \acrshort{psd} matrix can be equivalently defined as\footnote{The LHS of~\eqref{eq:cyc_psd_mat2} contains a 1-dimensional measure, whereas the RHS is a 2-dimensional correlation, hence, the number of $\der\sigma$ to enforce equality is different in both cases.}
\begin{subequations}\label{eq:cyc_psd_mat}
	\begin{align}
    	[\cycmat{X}(\sigma)]_{r,c}\der\sigma&=\expec\bigl[\der\nu^{(r)}_x(\sigma)\der\nu^{(c)*}_x(\sigma)\bigr]\\
	    [\cycmat{X}(\sigma)]_{r,c}&=\cyc_{X}\bigl(\tfrac{r-c}{P},\sigma+\tfrac{r}{P}\bigr)\der\sigma, \label{eq:cyc_psd_mat2}
	\end{align}
\end{subequations}
where $\der\nu_x^{(p)}(\sigma)\triangleq\der\nu_x(\sigma+\sfrac{p}{P})$.
This shows how $\cycmat{X}(\sigma)$ can be constructed from $P^2$ equally spaced elements of the two-dimensional spectral correlation $\cyc_{X}(\alpha,f)$, as illustrated in Fig.~\ref{fig:psd}.

The next result relates the \acrshort{cl} and \acrshort{kl} expansions of a \acrshort{cs} process.
\begin{corollary}{(Connection between \acrshort{cl} and \acrshort{kl} expansions of \acrshort{cs} processes)}\label{cor:cl-kl1}
    The \acrshort{kl} spectral representation of a \acrshort{cs} random process $\{x(n)\}$ is related to its \acrshort{cl} spectral representation through the following equality:
    \begin{equation}
        \widetilde{\mathbf{x}}(\sigma)=\herm{\mathbf{B}_X}(\sigma)\Breve{\mathbf{x}}(\sigma),\label{eq:prop_1}
    \end{equation}
    where
    \begin{equation}
    	\begin{aligned}
        	\widetilde{\mathbf{x}}(\sigma)&\triangleq\trans{\bigl[\der\xi^{(0)}_x(\sigma),\dots,\der\xi^{(P-1)}_{x}(\sigma)\bigr]}\\
        	\Breve{\mathbf{x}}(\sigma)&\triangleq\trans{\bigl[\der\nu^{(0)}_x(\sigma),\dots,\der\nu^{(P-1)}_{x}(\sigma)\bigr]}
	    \end{aligned}
    \end{equation}
	are constructed from $P$ samples of the \acrshort{kl} and \acrshort{cl} spectral processes, respectively, and {\normalfont $\der\xi^{(p)}_x(\sigma)\triangleq\der\xi_x(\sigma+\sfrac{p}{P})$}.
	Matrix {\normalfont $\mathbf{B}_X(\sigma)$} contains the eigenbasis of {\normalfont $\cycmat{X}(\sigma)$}:
    \begin{equation}
        \mathbf{B}_X(\sigma)\triangleq\bigl[\mathbf{b}^{(0)}_X(\sigma),\dots,\mathbf{b}^{(P-1)}_X(\sigma)\bigr].\label{eq:basis}
    \end{equation}
\end{corollary}
\begin{IEEEproof}
    See Appendix~\ref{app:cl-kl}.
\end{IEEEproof}

From this result, the \acrshort{kl} transform of a \acrshort{cs} process can be interpreted as the \acrshort{cl} transform of its (Gladyshev) harmonic representation series vector~\cite[Sec.~3.10]{Gardner2006}, followed by a decorrelation through unitary matrix $\mathbf{B}_X(\sigma)$.
A straightforward consequence of this is summarized in the following corollary.
\begin{corollary}{(Connection between cyclic \acrshort{psd} and \acrshort{kl}-\acrshort{psd} matrices of \acrshort{cs} processes)}\label{cor:cl-kl2}
    The \acrshort{kl}-\acrshort{psd} matrix
    {\normalfont
    \begin{equation}
	    \begin{aligned}
	        \klcycmat{\mathcal{X}}(\sigma)\der\sigma&\triangleq\mathrm{Diag}\bigl(\klcyc_{\mathcal{X}}^{(0)}(\sigma),\dots,\klcyc_{\mathcal{X}}^{(P-1)}(\sigma)\bigr)\der\sigma\\
	        &=\expec\bigl[\widetilde{\mathbf{x}}(\sigma)\herm{\widetilde{\mathbf{x}}}(\sigma)\bigr]
	    \end{aligned}\label{eq:cor_1}
    \end{equation}}
    contains the $P$ eigenvalues of the cyclic \acrshort{psd} matrix $\cycmat{X}(\sigma)$.
\end{corollary}
\begin{IEEEproof}
    Using relationships~\eqref{eq:cyc_psd_mat} and~\eqref{eq:prop_1}, we have:
    \begin{equation}
	    \begin{aligned}
        	\klcycmat{\mathcal{X}}(\sigma)\der\sigma&=\herm{\mathbf{B}_X}(\sigma)\expec\bigl[\Breve{\mathbf{x}}(\sigma)\herm{\Breve{\mathbf{x}}}(\sigma)\bigr]\mathbf{B}_X(\sigma)\\
	        &=\bigl(\herm{\mathbf{B}_X}\cycmat{X}\mathbf{B}_X\bigr)(\sigma)\der\sigma.
	    \end{aligned}
    \end{equation}
\end{IEEEproof}

\subsection{Temporal dependence of KL transforms}\label{ssec:temporal}

    While the reference time $n_0$ has been assumed 0 and omitted from the previous analysis, its impact on \acrshort{kl} representations should be addressed.
    In general, an arbitrary time shift will affect the \acrshort{kl} expansion of a random process.

    The two classes of stochastic processes considered here are special in this sense.
    The \acrshort{kl} basis of a \acrshort{wss} process remains unchanged by a temporal shift, whereas that of a \acrshort{cs} process depends on the reference time~\cite{Riba2022}:
    \begin{equation}
        \phi_{\mathrm{CS}}^{(p)}(n,\sigma,n_0)=\sum_{q=0}^{P-1}b_{X}^{(p)}(q,\sigma)\euler^{\imunit2\pi n_0\frac{q}{P}}\euler^{\imunit2\pi n(\sigma+\frac{q}{P})}.
    \end{equation}
    Regarding their spectral representation, a translation in time becomes a phase shift in the \acrshort{kl} domain for both of them:
    \begin{equation}
	    \begin{aligned}
	        \der\xi_x^{(p)}(\sigma,n_0)&\triangleq\kl\bigl\{x(n+n_0)\bigr\}(p,\sigma)\\
	        &=\kl\bigr\{x(n)\bigr\}(p,\sigma)\,\euler^{\imunit2\pi n_0\sigma}.
	    \end{aligned}\label{eq:time-shift}
    \end{equation}
    This property is derived for \acrshort{cs} processes in Appendix~\ref{app:time-shift}\footnote{Refer to~\cite[Eq.~(4.33)]{Wang2012} for the \acrshort{wss} case.}.

    On the contrary, the \acrshort{kl} spectrum of both classes is invariant to temporal translations, since the effect of the phase in~\eqref{eq:time-shift} disappears.
    This phenomenon is easily seen by using definition~\eqref{eq:kl-psd}:
    \begin{align}
		\klcyc_{\mathcal{X}}^{(p)}(\sigma,n_0)\der\sigma&=\expec\bigl[\bigl\lvert\der\xi_x^{(p)}(\sigma,n_0)\bigr\rvert^2\bigr]=\expec\bigl[\bigl\lvert\der\xi_x^{(p)}(\sigma)\euler^{\imunit2\pi 	n_0\sigma}\bigr\rvert^2\bigr]\nonumber\\
		&=\expec\bigl[\bigl\lvert\der\xi_x^{(p)}(\sigma)\bigr\rvert^2\bigr]\equiv\klcyc_{\mathcal{X}}^{(p)}(\sigma)\der\sigma.\label{eq:time_depend}
    \end{align}
    Since we are interested in obtaining asymptotic theoretical bounds, this property simplifies the subsequent derivations.
    
\subsection{Parseval's Theorem in the KL domain}\label{ssec:parseval}

    The average power of random process $\{x(n)\}$ is defined as
    \begin{equation}\label{eq:power_def}
        \power{x} \triangleq \smashoperator[l]{\lim_{N\to\infty}} \frac{1}{N} \smashoperator{\sum_{n=-\frac{N}{2}}^{\frac{N}{2}-1}} \expec\bigl[\lvert x(n)\rvert^2\bigr].
    \end{equation}
    Since the \acrshort{kl} transform is unitary, it preserves power in the spectral domain.
    A result similar to \textit{Parseval's theorem} for the frequency domain can thus be stated~\cite[Eq.~(2.80)]{Wang2012}:
	\begin{equation}
		\power{x} = \int_{\mathrlap{0}}^{\mathrlap{1}} \expec\bigl[\lvert\der\xi_x(\lambda)\rvert^2\bigr] = \smashoperator{\sum_{p=0}^{P-1}} \int_{\mathrlap{0}}^{\mathrlap{\sfrac{1}{P}}} \klcyc_{\mathcal{X}}^{(p)}(\sigma)\der\sigma.
	\end{equation}
    Using the segmented \acrshort{kl} spectrum allows to more compactly express the previous integral, by transforming the summation into a trace and applying Corollary~\ref{cor:cl-kl2}:
    \begin{equation}
	    \power{x} =  \int_{\mathrlap{0}}^{\mathrlap{\sfrac{1}{P}}} \trace\bigl[\klcycmat{\mathcal{X}}(\sigma)\bigr] \der\sigma = \int_{\mathrlap{0}}^{\mathrlap{\sfrac{1}{P}}} \trace\bigl[\cycmat{X}(\sigma)\bigr] \der\sigma.
    \end{equation}
    This way, we have obtained the power of a \acrshort{cs} process in terms of its cyclic \acrshort{psd} matrix.

\subsection{Energy compaction and minimum representation entropy}\label{ssec:optimal}

	While most unitary transforms exhibit some sort of energy concentration~\cite[Sec.~9.2.3]{Wang2012}, the \acrshort{kl} transform does it optimally.
	Let $\{\theta(n,\lambda)\}$ form an arbitrary unitary basis for a process $\{x(n)\}$:
	\begin{equation}
		x(n) = \int_{\mathrlap{0}}^{\mathrlap{1}} \theta(n,\lambda) \der\eta_x(\lambda),
	\end{equation}
	where $\der\eta_x(\lambda)$ is its spectral representation, \ie
	\begin{equation}
		\der\eta_x(\lambda) = \smashoperator[l]{\lim_{N\to\infty}} \frac{1}{N}  \smashoperator{\sum_{n=-\frac{N}{2}}^{\frac{N}{2}-1}} x(n) \theta^*(n,\lambda).
	\end{equation}
	Since we are dealing with a unitary transform, the power of $\{x(n)\}$ can be obtained in the spectral domain due to Parseval's theorem:
	\begin{equation}
		\power{x} = \int_{\mathrlap{0}}^{\mathrlap{1}} \expec\bigl[\lvert\der\eta_x(\lambda)\rvert^2\bigr] \triangleq \int_{\mathrlap{0}}^{\mathrlap{1}} \klcyc_{\mathcal{X},\theta}(\lambda) \der\lambda.
	\end{equation}
	We define the power contained in a portion $[0,\rho)$ of the spectrum:
	\begin{equation}\label{eq:fractional_power}
		\power{x}(\rho,\{\theta(n,\lambda)\}) \triangleq \int_{\mathrlap{0}}^{\mathrlap{\rho}} \klcyc_{\mathcal{X},\theta}(\lambda) \der\lambda,\quad0<\rho<1.
	\end{equation}
	Of all possible bases that fulfill~\eqref{eq:unitary}, we aim for the one that maximizes it for all $\rho$; \ie
	\begin{equation}
		\begin{aligned}
		\{\phi(n,\lambda)\} &= \argmax_{\{\theta(n,\lambda)\}} \power{x}(\rho,\{\theta(n,\lambda)\}) \\
		& \mathrm{s.t.} \quad \smashoperator{\sum_{n\in\integs}} \theta(n,\lambda) \theta^*(n,\lambda') = \delta(\lambda-\lambda').
	\end{aligned}
	\end{equation}
	The solution to this problem, which can be obtained through Lagrange multipliers~\cite[Th.~9.1]{Wang2012}, is the set of eigenfunctions $\{\theta(n,\lambda)\}$ that satisfies
	\begin{equation}
		\smashoperator{\sum_{l\in\integs}} \corr{x}(l,k-l) \theta(l,\lambda) = \chi(\lambda) \theta(k,\lambda).
	\end{equation}
	By definition (see~\eqref{eq:eigenequation}), this is the \acrshort{kl} basis $\{\phi(n,\lambda)\}$, and $\chi(\lambda)$ must be the \acrshort{kl}-\acrshort{psd} $\klcyc_{\mathcal{X}}(\lambda)$.
	
	This optimality in energy compaction can be quantified with a metric known as \textit{representation entropy}~\cite{Mitra2002} or \textit{spectral entropy}~\cite{Yang2005}, which indicates the compression of information achieved by a spectral representation.
	Let $\mathrm{p}_{x,\theta}(\lambda)\triangleq\klcyc_{\mathcal{X},\theta}(\lambda)/\power{x}$ be the normalized spectrum of $\{x(n)\}$ for a given orthonormal basis $\{\theta(n,\lambda)\}$.
	Since $\mathrm{p}_{x,\theta}(\lambda)$ is nonnegative and its integral is 1, it can be interpreted as a probability density function.
	We may then define the differential entropy associated with its corresponding representation as
	\begin{equation}
		\mathrm{h}(\{\theta(n,\lambda)\}) \triangleq -\int_{\mathrlap{0}}^{\mathrlap{1}} \mathrm{p}_{x,\theta}(\lambda) \cdot \ln(\mathrm{p}_{x,\theta}(\lambda))\der\lambda.
	\end{equation}
	\begin{proposition}{(Representation entropy)}\label{prop:entropy}
		The \acrshort{kl} expansion has the minimum representation entropy:
		\begin{equation}\label{eq:entropy}
			\mathrm{h}(\{\phi(n,\lambda)\}) \leq \mathrm{h}(\{\theta(n,\lambda)\}), \quad\forall\{\theta(n,\lambda)\}.
		\end{equation}
	\end{proposition}
	\begin{IEEEproof}
		Let $\mathrm{f}:[0,1]\mapsto\reals_{\geq0}$ be the continuous concave function $\mathrm{f}(a)\triangleq-a\cdot\ln(a)$.
		Knowing that
		\begin{equation}
			\power{x}(\rho,\{\phi(n,\lambda)\}) \geq \power{x}(\rho,\{\theta(n,\lambda)\}), \quad \forall\{\theta(n,\lambda)\},
		\end{equation}
		using~\cite[Def.~14.H.1]{Marshall2011} it can be stated that $\klcyc_{\mathcal{X},\theta}\prec\klcyc_{\mathcal{X},\phi}$.
		By~\cite[Prop.~14.H.1.a]{Marshall2011}, we can say that
		\begin{equation}
			\int_{\mathrlap{0}}^{\mathrlap{1}} \mathrm{f}\bigl(\mathrm{p}_{x,\phi}(\lambda)\bigr) \der\lambda \leq \int_{\mathrlap{0}}^{\mathrlap{1}} \mathrm{f}\bigl(\mathrm{p}_{x,\theta}(\lambda)\bigr)\der\lambda, \quad \forall\{\theta(n,\lambda)\},
		\end{equation}
		which is equivalent to~\eqref{eq:entropy}.
	\end{IEEEproof}
	
	Representation entropy is a measure of spectral redundancy, such as the one emerging from the spectrally correlated components of \acrshort{cs} processes.
	It is of interest to determine this form of frequency diversity~\cite[Sec.~9.2.3]{Schreier2010} since it is precisely the signal characteristic that synchronous processing exploits.

	\section{Cyclic Wiener filtering: problem statement}\label{sec:wiener}
		\noindent The previous formulation of \acrshort{cs} processes can be applied to the theory of Wiener filtering in order to obtain asymptotic bounds on the achievable performance.
We will focus on the following framework.
We deal with an observation $\{x(n)\}$ of a reference signal $\{d(n)\}$ distorted by additive noise $\{z(n)\}$:
\begin{equation}
    x(n)=d(n)+z(n).\label{eq:additive}
\end{equation}
They are assumed uncorrelated (\ie $\expec[d(n)z^*(n')]=0$, $\forall n,n'$).
Either $\{d(n)\}$ or $\{z(n)\}$ will be \acrshort{wss} with known \acrshort{psd}, while the other one will be \acrshort{cs}.
This combination of random processes presents many desirable properties.
By passing $\{x(n)\}$ through a stable \acrshort{lti} filter, its \acrshort{wss} component can be whitened~\cite[Sec.~11.3.3]{Oppenheim2017}, while the \acrshort{cs} properties of the other one are preserved, as proven in~\cite[Sec.~6.8.1]{Hurd2007}, as long as the full spectral support of the \acrshort{cs} component is within the filter band.
Therefore, and without loss of generality, the \acrshort{wss} process will be assumed white from this point onwards.

White \acrshort{wss} processes are of great interest because they admit any orthonormal \acrshort{kl} eigenbasis.
This allows the following simplification:
\begin{align}
	\der\xi_x^{(p)}(\sigma)&=\kl\bigl\{x(n)\bigr\}(p,\sigma)=\kl\bigl\{d(n)+z(n)\bigr\}(p,\sigma)\nonumber\\
	&=\kl\bigl\{d(n)\bigr\}(p,\sigma)+\kl\bigl\{z(n)\bigr\}(p,\sigma)\nonumber\\
	&=\der\xi_d^{(p)}(\sigma)+\der\xi_z^{(p)}(\sigma).
\end{align}
This simple model will provide various insights on the filtering problem and can be adapted to represent different cases of interest\footnote{A more realistic model would account for signal sampling being incommensurate with the true \acrshort{cs} period, requiring the use of a more general treatment~\cite[Sec.~1.3]{Napolitano2012}. It has been omitted from this work in favor of a clearer exposition.}:
\begin{itemize}
    \item \textbf{\acrshort{cs} signal $\{d(n)\}$ and white \acrshort{wss} noise $\{z(n)\}$:} this is a typical scenario in digital wireless communications, which usually involve the reception of \acrshort{cs} signals hindered by additive noise~\cite{Gardner1993}.
    \item \textbf{Wideband signal $\{d(n)\}$ and narrowband \acrshort{cs} interference $\{z(n)\}$:} in this scenario, which is often encountered in spread-spectrum systems, the high degree of predictability of $\{z(n)\}$ is used against the unpredictability of the desired term to improve signal separation.
    The nonstationarity of $\{d(n)\}$ is not exploited, and thus it is treated as \acrshort{wss}.
    Instead, the \acrshort{cs} nature of $\{z(n)\}$ can be harnessed beyond the spectral domain to further reduce the distortion of the desired component after a cancellation filter~\cite{Bershad2016}.
\end{itemize}
We will only develop results for the first case for brevity.
Extending the following analysis to the second one is straightforward and yields complementary expressions.

Our objective is to design a filter $\mathrm{w}\{\cdot\}$, such that it minimizes the average filtering error power, or \acrfull{mse}, for a time interval $N_1\leq n\leq N_2$:
\begin{equation}\label{eq:mse}
	\begin{aligned}
		\mse(N_1,N_2) &\triangleq \frac{1}{N_2-N_1} \smashoperator[r]{\sum_{n=N_1}^{N_2}} \expec\bigl[\lvert d(n)-\widehat{d}(n)\rvert^2\bigr] \\
		&= \frac{1}{N_2-N_1} \sum_{n}\mse(n).
	\end{aligned}
\end{equation}
The filtered signal is $\bigl\{\widehat{d}(n)\bigr\}\triangleq\mathrm{w}\{x(n)\}$ and $\bigl\{e(n)\triangleq d(n)-\widehat{d}(n)\bigr\}$ is the error process.
Three different problems will be studied in this setting~\cite[Sec.~1.2]{vanTrees2013}: \textit{smoothing}, \textit{filtering} and \textit{prediction}.

	\section{Asymptotic smoothing performance}\label{sec:noncausal}
		\noindent When future samples of process $\{x(n)\}$ are available, the previous problem is referred to as \textit{non-causal} or \textit{smoothing}~\cite[Sec.~9.1]{vanTrees2013}.
This case is common in applications without latency restrictions.
Since we are interested in the asymptotically achievable performance, the limit
\begin{equation}
	\mse_{\mathrm{nc}}\triangleq\smashoperator{\lim_{N\to\infty}}\mse\bigl(-\tfrac{N}{2},\tfrac{N}{2}-1\bigr)\label{eq:mse_nc}
\end{equation}
is considered.
Using Parseval's theorem, the error power can be expressed as
\begin{equation}\label{eq:err_power}
	\mse_{\mathrm{nc}} = \smashoperator{\sum_{p=0}^{P-1}} \int_{\mathrlap{0}}^{\mathrlap{\sfrac{1}{P}}} \expec\bigl[\bigl\lvert\der\xi_e^{(p)}(\sigma)\bigr\rvert^2\bigr],
\end{equation}
where $\{\der\xi_e^{(p)}(\sigma)\}$ is the \acrshort{kl} transform of the error process $\{e(n)\}$:
\begin{equation}
	\der\xi_e^{(p)}(\sigma)\triangleq\mathrm{KL}\bigl\{d(n)-\widehat{d}(n)\bigr\}(p,\sigma).
\end{equation}

For a more convenient analysis, and without loss of generality\footnote{This is ensured by the uncorrelatedness between spectral components provided by the \acrshort{kl} representation.}, we set the filter in the \acrshort{kl} domain to be a linear operator applied onto the observed signal $\{x(n)\}$:
\begin{equation}
	\der\xi_{\widehat{d}}^{(p)}(\sigma)\triangleq\mathcal{W}_p^*(\sigma)\cdot\der\xi_x^{(p)}(\sigma),\label{eq:transfer}
\end{equation}
where $\{\der\xi_{\widehat{d}}^{(p)}(\sigma)\}$ is the \acrshort{kl} representation of the filtered signal $\{\widehat{d}(n)\}$.
The previous expression can be seen as a natural extension of the transfer function in the frequency domain, commonly employed to deal with \acrshort{wss} processes through \acrshort{lti} systems: it relates every input to its corresponding output across the full spectrum.

Using~\eqref{eq:transfer}, we can express the spectral representation of $\{e(n)\}$ in terms of the ones of $\{d(n)\}$ and $\{\widehat{d}(n)\}$:
\begin{equation}
	\der\xi_e^{(p)}(\sigma)=\der\xi_d^{(p)}(\sigma)-\der\xi_{\widehat{d}}^{(p)}(\sigma).
\end{equation}
This decoupled structure is achieved by both signals sharing the same \acrshort{kl} basis, due to the noise component being white.
We can harness this simplicity to design filter $\mathcal{W}_p^*(\sigma)$, such that it minimizes the average error power at every point $(p,\sigma)$ of the spectrum:
\begin{equation}
	\mse_p(\sigma)\triangleq\expec\bigl[\bigl\lvert\der\xi_e^{(p)}(\sigma)\bigr\rvert^2\bigr].
\end{equation}
To obtain it, the derivative of each \acrshort{mse} term must be nullified:
\begin{gather}
	\begin{aligned}
		\frac{\der\mse_p(\sigma)}{\der\mathcal{W}_p^*(\sigma)}&=\mathcal{W}_p(\sigma)\cdot\expec\bigl[\bigl\lvert\der\xi_x^{(p)}(\sigma)\bigr\rvert^2\bigr]\\
		&\quad-\expec\bigl[\der\xi_d^{(p)*}(\sigma)\der\xi_x^{(p)}(\sigma)\bigr]\equiv0
	\end{aligned}\\
	\mathcal{W}_p(\sigma)=\frac{\expec\bigl[\der\xi_d^{(p)*}(\sigma)\der\xi_x^{(p)}(\sigma)\bigr]}{\expec\bigl[\bigl\lvert\der\xi_x^{(p)}(\sigma)\bigr\rvert^2\bigr]}\triangleq\frac{\klcyc_{\mathcal{DX}}^{(p)*}(\sigma)\der\sigma}{\klcyc_{\mathcal{X}}^{(p)}(\sigma)\der\sigma}.\label{eq:non-causal}
\end{gather}
This solution is the \acrshort{kl} extension of the classical non-causal Wiener filter in the frequency domain~\cite[Eq.~(8.144)]{vanTrees2013}.

\subsection{Non-causal MMSE and coherence}
	Plugging~\eqref{eq:non-causal} back into~\eqref{eq:err_power} produces the \acrshort{mmse}\footnote{Notice how~\eqref{eq:wiener-error} resembles~\cite[Eq.~(13.115)]{Gardner1986}.}:
	\begin{equation}
		\begin{aligned}
			\mmse_{\mathrm{nc}} &= \smashoperator{\sum_{p=0}^{P-1}} \int_{\mathrlap{0}}^{\mathrlap{\sfrac{1}{P}}} \klcyc_{\mathcal{D}}^{(p)}(\sigma)\bigl(1-\lvert\gamma_p(\sigma)\rvert^2\bigr)\der\sigma,
		\end{aligned}\label{eq:wiener-error}
	\end{equation}
	where
	\begin{equation}
		\lvert\gamma_p(\sigma)\rvert^2\triangleq\frac{\bigl\lvert\klcyc_{\mathcal{DX}}^{(p)}(\sigma)\bigr\rvert^2}{\klcyc_{\mathcal{X}}^{(p)}(\sigma)\klcyc_{\mathcal{D}}^{(p)}(\sigma)}\in[0,1]\label{eq:spec_coh}
	\end{equation}
	is the \acrshort{kl} version of the \textit{squared spectral coherence} in the frequency domain~\cite[Eq.~(3.6)]{Ramirez2022}.
	This is a measure of linear dependency between observation $\der\xi_x^{(p)}(\sigma)$ and reference $\der\xi_d^{(p)}(\sigma)$.
	Its value is directly related to the \acrshort{mmse}: as clearly stated in~\eqref{eq:wiener-error}, the closer $|\gamma_p(\sigma)|^2$ is to 1 for all $(p,\sigma)$, the lower the achievable \acrshort{mmse} will be.
	
	Representing~\eqref{eq:wiener-error} in matrix form will provide valuable insights on the filtering problem.
	Let
	\begin{equation}
		\klcycmat{\mathcal{DX}}(\sigma)\triangleq\Diag\bigl(\klcyc^{(0)}_{\mathcal{DX}}(\sigma),\dots,\klcyc^{(P-1)}_{\mathcal{DX}}(\sigma)\bigr)
	\end{equation}
	be the \acrshort{kl} correlation matrix between $\{d(n)\}$ and $\{x(n)\}$.
	Then, the summation in~\eqref{eq:wiener-error} can be expressed as a trace:
	\begin{align}
		&\mmse_{\mathrm{nc}}=\\
		&\quad\int_{\mathrlap{0}}^{\mathrlap{\sfrac{1}{P}}} \trace\bigl[\klcycmat{\mathcal{D}}(\sigma)\bigl(\id{P}-\bigl(\klcycmat{\mathcal{D}}^{-1}\klcycmat{\mathcal{DX}}\klcycmat{\mathcal{X}}^{-1}\herm{\klcycmat{\mathcal{DX}}}\bigr)(\sigma)\bigr)\bigr]\der\sigma.\nonumber
	\end{align}
	Using the circularity property of the trace operator together with decomposition
	\begin{equation}
		\cycmat{DX}(\sigma)\triangleq\bigl(\mathbf{B}_D\klcycmat{\mathcal{DX}}\herm{\mathbf{B}_X}\bigr)(\sigma),
	\end{equation}
	which is a straightforward consequence of Corollary~\ref{cor:cl-kl2}, we can express the non-causal \acrshort{mmse} in terms of cyclic \acrshort{psd} matrices\footnote{The matrix inside the trace in~\eqref{eq:mse-cyclic} is the \textit{error covariance matrix} in~\cite[Sec.~3.6]{Ramirez2022}.}:
	\begin{equation}
		\mmse_{\mathrm{nc}} = \int_{\mathrlap{0}}^{\mathrlap{\sfrac{1}{P}}} \trace\bigl[\cycmat{D}(\sigma)\bigl(\id{P}-\coherence_{DX}(\sigma)\herm{\coherence_{DX}}(\sigma)\bigr)\bigr]\der\sigma,\label{eq:mse-cyclic}
	\end{equation}
	where
	\begin{equation}
		\coherence_{DX}(\sigma)\triangleq\bigl(\cycmat{D}^{-\sfrac{1}{2}}\cycmat{DX}\cycmat{X}^{-\sfrac{1}{2}}\bigr)(\sigma),\label{eq:coherence}
	\end{equation}
	is the \textit{coherence matrix}~\cite[Def.~3.3]{Ramirez2022}.
	It generalizes the notion of spectral coherence presented in~\eqref{eq:spec_coh}: the more $\coherence_{DX}(\sigma)$ resembles a unitary matrix for all $\sigma$, the lower the \acrshort{mmse} will be.
	This mathematical structure is very recurrent in the literature and is strongly related to canonical correlations and principal angles~\cite[Ch.~3]{Ramirez2022}.
	
	On a final note, we can particularize the non-causal \acrshort{mmse}~\eqref{eq:wiener-error} for our additive model, in which the white noise has \acrshort{kl}-\acrshort{psd} $\klcyc_{\mathcal{Z}}^{(p)}(\sigma)=\power{z}$, for all $(p,\sigma)$.
	The filter expression~\eqref{eq:non-causal} simplifies to
	\begin{equation}
		\mathcal{W}_p(\sigma)=\frac{\klcyc_{\mathcal{D}}^{(p)}(\sigma)}{\klcyc_{\mathcal{D}}^{(p)}(\sigma)+\power{z}},
	\end{equation}
	with which we obtain
	\begin{equation}
		\begin{aligned}
			\mmse_{\mathrm{nc}}&=\sum_{p=0}^{P-1}\int_{\mathrlap{0}}^{\mathrlap{\sfrac{1}{P}}} \der\sigma \frac{\klcyc_{\mathcal{D}}^{(p)}(\sigma)\power{z}}{\klcyc_{\mathcal{D}}^{(p)}(\sigma)+\power{z}}\\
			&=\power{z} \int_{\mathrlap{0}}^{\mathrlap{\sfrac{1}{P}}} \trace\bigl[\cycmat{D}(\sigma)(\cycmat{D}(\sigma)+\power{z}\id{P})^{-1}\bigr]\der\sigma.
		\end{aligned}\label{eq:mmse-simple}
	\end{equation}

\subsection{Equivalence between KL Wiener filter and CWF}
	It is known that the \acrshort{cwf}, which is the optimum filter for \acrshort{cs} signals, is periodically time variant~\cite{Gardner1993}.
	It is usually implemented through a \acrshort{fresh} structure, which has the following formulation in the frequency domain~\cite[Sec.~10.3]{Schreier2010}:
	\begin{equation}
		\begin{aligned}
			\der\nu_{\widehat{d}}^{(p)}(\sigma)&\triangleq\sum_{q=0}^{P-1}\mathrm{W}_q^*(\sigma+\tfrac{p}{P})\der\nu_x^{(q)}(\sigma+\tfrac{p}{P})\\
			&\triangleq\herm{\Breve{\mathbf{w}}}_p(\sigma)\mathbf{P}_{\Pi}^{p}\Breve{\mathbf{x}}(\sigma),
		\end{aligned}\label{eq:cyclic_wiener}
	\end{equation}
	where $\mathbf{P}_{\Pi}$ is a \textit{cyclic permutation matrix}~\cite[Eq.~(4.1.1)]{Bernstein2018}:
	\begin{equation}
		\mathbf{P}_{\Pi}\triangleq\begin{bmatrix}
			\mathbf{0}_{(P-1)} & \id{(P-1)}\\
			1 & \trans{\mathbf{0}_{(P-1)}}
		\end{bmatrix}.
	\end{equation}
	Applying Parseval's theorem onto~\eqref{eq:mse_nc}, we may express the (non-causal) error power in the \acrshort{cl} domain:
	\begin{equation}
		\mse_{\mathrm{nc}}=\sum_{p=0}^{P-1}\int_{\mathrlap{0}}^{\mathrlap{\sfrac{1}{P}}} \expec\bigl[\bigl\lvert\der\nu_{d}^{(p)}(\sigma)-\herm{\Breve{\mathbf{w}}}_p(\sigma)\mathbf{P}_{\Pi}^p\Breve{\mathbf{x}}(\sigma)\bigr\rvert^2\bigr].\label{eq:mse-cwf}
	\end{equation}
	
	To obtain the optimal \acrshort{fresh} filter, we must nullify the derivative of the previous expression with respect to $\Breve{\mathbf{w}}^*_p(\sigma)$:
	\begin{align}
		\mathbf{\nabla}_{\Breve{\mathbf{w}}^*_p(\sigma)}\mse_{\mathrm{nc}}^{(p)}(\sigma)&=\mathbf{P}_{\Pi}^{p}\bigl(\cycmat{X}(\sigma)\mathbf{P}_{\Pi}^{-p}\Breve{\mathbf{w}}_p(\sigma)-\mathbf{s}_{XD}^{(p)}(\sigma)\bigr)\equiv0\nonumber\\
		\Breve{\mathbf{w}}_p(\sigma)&=\mathbf{P}_{\Pi}^{p}\cycmat{X}^{-1}(\sigma)\mathbf{s}_{XD}^{(p)}(\sigma),\label{eq:cyclic_wiener_filter}
	\end{align}
	where
	\begin{equation}
		\mathbf{s}_{XD}^{(p)}(\sigma)\der\sigma\triangleq\expec\bigl[\Breve{\mathbf{x}}(\sigma)\der\nu_d^{(p)*}(\sigma)\bigr]
	\end{equation}
	is the $p$th column of $\herm{\cycmat{DX}}(\sigma)\der\sigma$.
	The following result relates the \acrshort{kl} Wiener filter derived in~\eqref{eq:non-causal} with the \acrshort{cwf} obtained herein.
	
	\begin{theorem}{(\acrshort{cwf} in the \acrshort{kl} domain)}\label{thm:mult_cyclic}
		The \acrshort{cwf} is equivalent to the \acrshort{kl} Wiener filter.
	\end{theorem}
	\begin{IEEEproof}
		The proof is based on comparing the output signals obtained from the cyclic and \acrshort{kl} Wiener filters.
		Regarding the former, we simply plug~\eqref{eq:cyclic_wiener_filter} into~\eqref{eq:cyclic_wiener}:
		\begin{equation}
			\der\nu_{\widehat{d}}^{(p)}(\sigma)=\bigl(\mathbf{s}_{XD}^{(p)\mathrm{H}}\cycmat{X}^{-1}\Breve{\mathbf{x}}\bigr)(\sigma).
		\end{equation}
		As for the latter, we apply~\eqref{eq:non-causal} to~\eqref{eq:transfer}:
		\begin{equation}
			\der\xi_{\widehat{d}}^{(p)}(\sigma)=\frac{\klcyc_{\mathcal{DX}}^{(p)}(\sigma)}{\klcyc_{\mathcal{X}}^{(p)}(\sigma)}\der\xi_x^{(p)}(\sigma).\label{eq:kl_filtered}
		\end{equation}
		
		To compare them, we must express both filtered signals in the same domain.
		We achieve this by stacking $P$ samples of $\der\nu_{\widehat{d}}^{(p)}(\sigma)$ as
		\begin{equation}
			\begin{aligned}
				\breve{\hat{\mathbf{d}}}(\sigma)&\triangleq\trans{\bigl[\der\nu_{\widehat{d}}^{(0)}(\sigma),\dots,\der\nu_{\widehat{d}}^{(P-1)}(\sigma)\bigr]}\\
				&=(\cycmat{DX}\cycmat{X}^{-1}\Breve{\mathbf{x}})(\sigma),
			\end{aligned}
		\end{equation}
		and using Corollaries~\ref{cor:cl-kl1} and~\ref{cor:cl-kl2}:
		\begin{equation}
			\begin{aligned}
				\herm{\mathbf{B}}_D(\sigma)\breve{\hat{\mathbf{d}}}(\sigma)&=(\herm{\mathbf{B}}_D\cycmat{DX}\mathbf{B}_X\klcycmat{\mathcal{X}}^{-1}\herm{\mathbf{B}}_X\Breve{\mathbf{x}})(\sigma)\\
				&=(\klcycmat{\mathcal{DX}}\klcycmat{\mathcal{X}}^{-1}\widetilde{\mathbf{x}})(\sigma).
			\end{aligned}
		\end{equation}
		Since the $p$th term of the transformed vector corresponds to~\eqref{eq:kl_filtered}, we can assert that both expansions correspond to the same process, making the cyclic and linear \acrshort{kl} Wiener filters equivalent.
		This completes the proof.
	\end{IEEEproof}
	
	This result bears remarkable implications in the context of this paper.
	By studying a linear filtering problem in the \acrshort{kl} domain, we have accessed asymptotic performance results of periodically time variant filtering (namely~\eqref{eq:wiener-error}) without having to explicitly derive its expressions nor deal with the \acrshort{fresh} structure.

\subsection{Second order characterization of the error process}\label{ssec:sync_gain:second_order}

	The error process resulting after filtering~\eqref{eq:additive} with the \acrshort{cwf} has the following spectral representation:
	\begin{align}
		\breve{\mathbf{e}}(\sigma)&\triangleq\trans{\bigl[\der\nu_e^{(0)}(\sigma),\dots,\der\nu_e^{(P-1)}(\sigma)\bigr]}\\
		&=\Breve{\mathbf{d}}(\sigma)-\cycmat{D}(\sigma)(\cycmat{D}(\sigma)+\power{z}\id{P})^{-1}\bigl(\Breve{\mathbf{d}}(\sigma)+\Breve{\mathbf{z}}(\sigma)\bigr)\nonumber\\
		&=(\cycmat{D}(\sigma)+\power{z}\id{P})^{-1}\bigl(\power{z}\Breve{\mathbf{d}}(\sigma)-\cycmat{D}(\sigma)\Breve{\mathbf{z}}(\sigma)\bigr).\nonumber
	\end{align}
	Notice we have used the same notation as in Section~\ref{sec:kl}.
	Its cyclic spectrum matrix is
	\begin{equation}
		\cycmat{E}(\sigma)\der\sigma=\expec[\Breve{\mathbf{e}}(\sigma)\herm{\Breve{\mathbf{e}}}(\sigma)]
		=(\power{z}^{-1}\id{P}+\cycmat{D}^{-1}(\sigma))^{-1}\der\sigma.
	\end{equation}
	Since this matrix is not diagonal for $\power{z}>0$, it implies the error process of the \acrshort{cwf} is \acrshort{cs}.

\subsection{Synchronous gain}\label{ssec:sync_gain:sync_gain}

	To complement this analysis, it is of interest to define the \textit{synchronous gain}: the improvement in \acrshort{mse} that can be harnessed by taking cyclostationarity into account.
	Returning to the additive model~\eqref{eq:additive} with \acrshort{cs} reference, recall the \acrshort{mmse} expression in the Fourier domain~\eqref{eq:mse-cwf}.
	If $\{d(n)\}$ is erroneously treated as \acrshort{wss}, the classical non-causal Wiener filter~\cite[Eq.~(8.149)]{vanTrees2013} can be employed, which yields
	\begin{equation}\label{eq:mse-wss}
		\mmse_{\mathrm{nc,WSS}} = \smashoperator{\sum_{p=0}^{P-1}} \int_{0}^{\mathrlap{\sfrac{1}{P}}} \der\sigma \frac{\cyc_{D}\bigl(\sigma+\tfrac{p}{P}\bigr)\power{z}}{\cyc_{D}\bigl(\sigma+\tfrac{p}{P}\bigr)+\power{z}} ,
	\end{equation}
	as stated in~\cite[Eq.~(106)]{Guo2005}.
	Notice how~\eqref{eq:mse-wss}, which is a function of the diagonal elements of $\cycmat{D}(\sigma)$, is related to~\eqref{eq:mmse-simple}, which depends on its eigenvalues instead.
	With the two expressions, the synchronous gain is defined as
	\begin{equation}
		\zeta_{\mathrm{nc}} \triangleq \frac{\mmse_{\mathrm{nc}}}{\mmse_{\mathrm{nc,WSS}}}.
	\end{equation}
	
	One might expect this value to be less than 1, implying a reduction in \acrshort{mse} achieved by a more nuanced processing.
	This intuition can be formally proved using the theory of majorization~\cite{Marshall2011}, as in Proposition~\ref{prop:entropy}.
	We know that the eigenvalues of a Hermitian matrix majorize its diagonal elements~\cite[Th.~9.B.1.]{Marshall2011}, \ie $\diag\bigl(\cycmat{D}(\sigma)\bigr)\prec\diag\bigl(\klcycmat{\mathcal{D}}(\sigma)\bigr)$.
	Therefore, by~\cite[Prop.~4.B.1.]{Marshall2011}, we can assert that $\mmse_{\mathrm{nc}}\leq\mmse_{\mathrm{nc,WSS}}$.
	This ensures that there is a \acrshort{mse} improvement in performing a synchronous processing, compared to the conventional Wiener filter treatment.

	\section{Asymptotic filtering performance}\label{sec:causal}
		\noindent Non-causal filters are often referred to as \textit{unrealizable}~\cite[Sec.~4.1]{Proakis2013} because it may be unrealistic to assume the availability of future samples from a process.
For this reason, the results obtained in the previous section are to be understood as bounds on how accurate synchronous filtering can get, rather than an achievable performance in practice.

We define the \acrfull{snr} as the ratio between the average power of $\{d(n)\}$ and $\{z(n)\}$.
If we assume $\{d(n)\}$ has unit power, it reduces to $\snr\triangleq1/\power{z}$.
We may express the \acrshort{mmse}~\eqref{eq:mmse-simple} obtained from the non-causal \acrshort{cwf} in terms of this \acrshort{snr}:
\begin{equation}\label{eq:mmse_nc_snr}
	\mmse_{\mathrm{nc}}(\snr) = \smashoperator{\sum_{p=0}^{P-1}} \int_{\mathrlap{0}}^{\mathrlap{\sfrac{1}{P}}} \der\sigma \frac{\klcyc_{\mathcal{D}}^{(p)}(\sigma)}{\snr\cdot\klcyc_{\mathcal{D}}^{(p)}(\sigma)+1}.
\end{equation}

Let $\{x(n)\}_{n_a}^{n_b}$ be a sequence of consecutive samples from $\{x(n)\}$, for $n_a\leq n\leq n_b$.
The infinite past \textit{causal \acrshort{mmse}} at any $n_b$ given $\{x(n)\}_{n_a}^{n_b}$ is known to be~\cite[Sec.~4.2.1]{Levy2008}
\begin{multline}
	\mathrm{mmse}_{\mathrm{c}}(n_b,\snr)\triangleq\\
	\expec\bigl[\bigl\lvert d(n_b)-\expec\bigl[d(n_b)\vert\{x(n)\}_{-\infty}^{n_b};\snr\bigr]\bigr\rvert^2\bigr],
\end{multline}
and its time average is
\begin{equation}\label{eq:mmse_average}
	\mmse_{\mathrm{c}}(\snr) \triangleq \smashoperator[l]{\lim_{N\to\infty}} \tfrac{1}{N} \cdot \smashoperator{\sum_{n=n_b-N+1}^{n_b}}  \mathrm{mmse}_{\mathrm{c}}(n,\snr).
\end{equation}
Its derivation can be circumvented by employing the following fundamental result in nonlinear filtering.
\begin{theorem}{(Guo-Shamai-Verdú~\cite[Th.~8]{Guo2005})}\label{thm:gsv}
	The average (per unit time) \acrshort{mmse} for filtering can be obtained by averaging the smoothing \acrshort{mmse} over all possible values of \acrshort{snr}:
	\begin{equation}
		\mmse_{\mathrm{c}}(\snr)=\expec\bigl[\mmse_{\mathrm{nc}}(\Gamma)\bigr],\label{eq:gsv}
	\end{equation}
	where $\Gamma$ is distributed uniformly in the interval $[0,\snr]$.
\end{theorem}

In Section~\ref{ssec:sync_gain:second_order}, we have proved the error process after smoothing signal~\eqref{eq:additive} with the \acrshort{cwf} is \acrshort{cs}.
Hence,~\eqref{eq:mmse_nc_snr} is valid for the (asymptotic) non-causal \acrshort{mmse} averaged over a single period $P$.
Since Theorem~\ref{thm:gsv} links it with its causal counterpart, this implies $\mmse_{\mathrm{c}}(\snr)$ averaged over a single period is equivalent to~\eqref{eq:mmse_average}, which is averaged over the full infinite past.

To obtain~\eqref{eq:gsv} from~\eqref{eq:mmse_nc_snr}, we perform the following expectation:
\begin{equation}\label{eq:causal_mmse}
	\begin{aligned}
		& \mmse_{\mathrm{c}}(\snr) = \frac{1}{\snr} \int_{\mathrlap{0}}^{\mathrlap{\snr}} \mmse_{\mathrm{nc}}(\Gamma) \der\Gamma\\
		&\quad = \frac{1}{\snr} \smashoperator{\sum_{p=0}^{P-1}} \int_{0}^{\mathrlap{\sfrac{1}{P}}} \der\sigma \int_{\mathrlap{0}}^{\mathrlap{\snr}} \der\Gamma \frac{\klcyc_{\mathcal{D}}^{(p)}(\sigma)}{\Gamma\cdot\klcyc_{\mathcal{D}}^{(p)}(\sigma)+1} \\
		&\quad = \frac{1}{\snr} \int_{0}^{\mathrlap{\sfrac{1}{P}}} \der\sigma \smashoperator{\sum_{p=0}^{P-1}} \ln\bigl(\snr\cdot\klcyc_{\mathcal{D}}^{(p)}(\sigma)+1\bigr) .
	\end{aligned}
\end{equation}
Transforming the sum of logarithms into a log-determinant, and making use of Corollary~\ref{cor:cl-kl2}, we can express this result in terms of the cyclic \acrshort{psd} and the coherence matrix~\eqref{eq:coherence}:
\begin{equation}
	\begin{aligned}
		&\mmse_{\mathrm{c}}(\snr)=\frac{1}{\snr}\int_{\mathrlap{0}}^{\mathrlap{\sfrac{1}{P}}} \ln\lvert\snr\cdot\klcycmat{\mathcal{D}}(\sigma)+\id{P}\rvert\der\sigma\\
		&\quad=\frac{1}{\snr}\int_{\mathrlap{0}}^{\mathrlap{\sfrac{1}{P}}} \ln\lvert\snr\cdot\cycmat{D}(\sigma)+\id{P}\rvert\der\sigma\\
		&\quad=\frac{-1}{\snr}\int_{\mathrlap{0}}^{\mathrlap{\sfrac{1}{P}}} \ln\bigl\lvert\id{P}-\coherence_{DX}(\sigma)\herm{\coherence_{DX}}(\sigma)\bigr\rvert\der\sigma.
	\end{aligned}
\end{equation}
This formula is related to various problems that require the detection of correlation and cyclostationarity~\cite[Ch.~8]{Ramirez2022}.
It also provides a link between \acrshort{mmse}, the coherence matrix and mutual information, since it involves a generalization of the \textit{information-theoretic coherence} from~\cite[Sec.~11.4]{Ramirez2022}.

\subsection{Synchronous gain}

	The causal \acrshort{mmse} assuming $\{d(n)\}$ is \acrshort{wss} is known to be~\cite[Eq.~(8.177)]{vanTrees2013}
	\begin{multline}
		\mmse_{\mathrm{c},\mathrm{WSS}}(\snr)=\\
		\sum_{p=0}^{P-1}\int_0^{\frac{1}{P}}\frac{\ln\bigl(\snr\cdot\cyc_{D}\bigl(\sigma+\tfrac{p}{P}\bigr)+1\bigr)}{\snr}\der\sigma.
	\end{multline}
	Using the same rationale based on majorization theory as in Section~\ref{ssec:sync_gain:sync_gain}, we once again state that the \acrshort{mse} can be reduced by synchronous processing, \ie
	\begin{equation}
		\zeta_{\mathrm{c}}\triangleq\frac{\mmse_{\mathrm{c}}}{\mmse_{\mathrm{c},\mathrm{WSS}}}\leq1.
	\end{equation}

\subsection{High SNR regime}\label{ssec:sync_gain:high_snr}

	Suppose that $\klcyc_{\mathcal{D}}(\lambda)$ in~\eqref{eq:causal_mmse} is concentrated inside a band $[0,B)$, for $0<B<1$, such that it is null outside of it:
	\begin{equation}
		\mmse_{\mathrm{c}}(\snr)=\int_{\mathrlap{0}}^{\mathrlap{B}}\der\lambda\frac{\ln\bigl(\klcyc_{\mathcal{D}}(\lambda)+\frac{1}{\snr}\bigr)}{\snr}+B\cdot\frac{\ln\snr}{\snr}.
	\end{equation}
	In the limit of $\snr\to\infty$, the causal \acrshort{mmse} decays as\footnote{We have assumed that $\ln\klcyc_{\mathcal{D}}(\lambda)$ is Riemann-integrable in $[0,B)$~\cite[Def.~9.5.1]{Choudary2014}, which is closely related to the \textit{regularity condition} in~\cite[Th.~4.3]{Doob1990}.}	
	\begin{align}
		\smashoperator{\lim_{\snr\to\infty}} \mmse_{\mathrm{c}}(\snr) &= \smashoperator{\lim_{\snr\to\infty}} \frac{\int_{0}^{B}\ln\klcyc_{\mathcal{D}}(\lambda)\der\lambda}{\snr} + B\cdot\frac{\ln\snr}{\snr} \nonumber\\
		& = \smashoperator{\lim_{\snr\to\infty}} B\cdot\frac{\ln\snr}{\snr} + O\bigl(\snr^{-1}\bigr).\label{eq:high_snr_causal}
	\end{align}
	This implies that, the more concentrated the \acrshort{kl} spectrum of $\{d(n)\}$ is (\ie the smaller $B$ is), the lower $\mmse_{\mathrm{c}}$ will be at high \acrshort{snr}.
	The same idea applies to its non-causal counterpart; it can be clearly seen by observing the limit of~\eqref{eq:mmse_nc_snr} at high \acrshort{snr} under mild conditions:
	\begin{align}\label{eq:high_snr_non-causal}
		\begin{aligned}
			\smashoperator{\lim_{\snr\to\infty}} \mmse_{\mathrm{nc}}(\snr) &= \smashoperator[l]{\lim_{\snr\to\infty}} \int_{\mathrlap{0}}^{\mathrlap{B}} \der\lambda \frac{\klcyc_{\mathcal{D}}(\lambda)}{\snr\cdot\klcyc_{\mathcal{D}}(\lambda)+1} \\
			&= \smashoperator{\lim_{\snr\to\infty}} B/\snr.
		\end{aligned}
	\end{align}
	From~\eqref{eq:high_snr_causal} and~\eqref{eq:high_snr_non-causal} we conclude that the causal \acrshort{mmse} asymptotically approaches the rate of decay of the non-causal \acrshort{mmse}, which is hyperbolic.

	\section{Asymptotic prediction performance}\label{sec:prediction}
		\noindent The final problem that will be analyzed is the one-step linear prediction of $\{x(n)\}$ based on $N$ past samples:
\begin{equation}
	\widehat{x}_N(n)=\herm{\mathbf{w}_N}(n)\mathbf{x}_N(n-1),
\end{equation}
where
\begin{equation}
	\mathbf{x}_N(n-1)\triangleq\trans{[x(n-N),\dots,x(n-2),x(n-1)]}.
\end{equation}
The \acrshort{mmse} in this case is given by
\begin{equation}
	\mmse_{\mathrm{p}}^{(N)}(n)\triangleq\smashoperator{\min_{\mathbf{w}_N(n)}}\expec\bigl[\lvert x(n)-\widehat{x}_N(n)\rvert^2\bigr].
\end{equation}
This error is achieved by the cyclic Wiener predictor, which is periodic and implements a synchronous processing.
It is known to be~\cite[Eq.~(2.10)]{Vaidyanathan2007}
\begin{multline}
	\mmse_{\mathrm{p}}^{(N)}(n)\\
	=\corr{x}(n,0)-\mathbf{r}_{x}^{(N)\mathrm{H}}(n)\bigl(\corrmat_{x}^{(N)}(n-1)\bigr)^{-1}\mathbf{r}_{x}^{(N)}(n),
\end{multline}
where the autocorrelation matrix and vector are respectively defined as
\begin{equation}
	\begin{aligned}
		\corrmat_x^{(N)}(n-1)&\triangleq\expec[\mathbf{x}_N(n-1)\herm{\mathbf{x}}_N(n-1)]\\
		\mathbf{r}_{x}^{(N)}(n)&\triangleq\expec[\mathbf{x}_N(n-1)x^*(n)].
	\end{aligned}
\end{equation}
Notice that the correlation matrix of size $N+1$ at instant $n$ can be expressed blockwise as
\begin{equation}
	\corrmat_{x}^{(N+1)}(n)=\begin{bmatrix}
		\corrmat_{x}^{(N)}(n-1) & \mathbf{r}_{x}^{(N)}(n)\\
		\mathbf{r}_{x}^{(N)\mathrm{H}}(n) & \corr{x}(n,0)
	\end{bmatrix}.
\end{equation}
Using the determinant formula for block matrices, we can express the \acrshort{mmse} as follows~\cite[Eq.~(2.41)]{Vaidyanathan2007}:
\begin{equation}\label{eq:determinants}
	\mmse_{\mathrm{p}}^{(N)}(n) = \frac{\bigl\lvert\corrmat_{x}^{(N+1)}(n)\bigr\rvert}{\bigl\lvert\corrmat_{x}^{(N)}(n-1)\bigr\rvert}.
\end{equation}

We are interested in the behavior of this error as past samples grow without bound, \ie
\begin{equation}
	\mmse_{\mathrm{p}}\triangleq\smashoperator{\lim_{N\to\infty}}\mmse_{\mathrm{p}}^{(N)}(n).
\end{equation}
\begin{theorem}{(Riba \& Vilà-Insa~\cite[Th.~2]{Riba2022})}\label{thm:pred}
	A lower bound on the \acrshort{mmse} of one-step prediction of a cyclostationary signal is given by
	\begin{equation}
		\mmse_{\mathrm{p}}=\exp\int_{\mathrlap{0}}^{\mathrlap{\sfrac{1}{P}}} \ln\lvert\cycmat{X}(\sigma)\rvert\der\sigma.
	\end{equation}
\end{theorem}
\begin{IEEEproof}
	See Appendix~\ref{app:pred}.
\end{IEEEproof}

A metric that is closely related to the \acrshort{mmse} in prediction is the \textit{spectral flatness} of \acrshort{wss} processes~\cite[Def.~6.1]{Vaidyanathan2007}, which we extend to \acrshort{cs} ones by defining the \textit{\acrshort{kl} spectral flatness}:
\begin{equation}
	\omega_x^2\triangleq\frac{\mmse_{\mathrm{p}}}{\power{x}}\in[0,1].
\end{equation}
This ratio measures the shape (\ie flatness) of the \acrshort{kl} spectrum of $\{x(n)\}$ and is directly linked to its representation entropy and predictability through synchronous processing~\cite[Sec.~6.6.1]{Vaidyanathan2007}.

\subsection{Synchronous gain}
	The minimum \acrshort{mmse} achievable from classical Wiener prediction is given by the \textit{Kolmogorov-Szegö theorem}~\cite[Eq.~(6.23)]{Vaidyanathan2007}:
	\begin{equation}
		\mmse_{\mathrm{p},\mathrm{WSS}}=\exp\int_{\mathrlap{0}}^{\mathrlap{\sfrac{1}{P}}} \ln\bigl\lvert\overline{\mathbf{S}}_{X}(\sigma)\overline{\mathbf{S}}\herm{\vphantom{\mathbf{S}}}_{X}(\sigma)\bigr\rvert\der\sigma.
	\end{equation}
	where $\overline{\mathbf{S}}_{X}(\sigma)\triangleq(\id{P}\circ\cycmat{X}(\sigma))^{\sfrac{1}{2}}$.
	In this case, the synchronous gain
	\begin{equation}
		\zeta_{\mathrm{p}}\triangleq\frac{\mmse_{\mathrm{p}}}{\mmse_{\mathrm{p},\mathrm{WSS}}}\leq1
	\end{equation}
	can be expressed compactly as a function of the \textit{spectral coherence matrix}~\cite[Sec.~8.2]{Ramirez2022}:
	\begin{equation}
		\overline{\mathbf{C}}_{X}(\sigma)\triangleq(\overline{\mathbf{S}}\vphantom{\mathbf{S}}^{-1}_{X}\cycmat{X}\overline{\mathbf{S}}\vphantom{\mathbf{S}}^{-\mathrm{H}}_{X})(\sigma).
	\end{equation}
	This fact was stated in~\cite{Riba2022}:
	\begin{equation}
		\begin{aligned}
			\zeta_{\mathrm{p}}&=\exp\int_{\mathrlap{0}}^{\mathrlap{\sfrac{1}{P}}} \ln\lvert\cycmat{X}(\sigma)\rvert-\ln\bigl\lvert\overline{\mathbf{S}}_{X}(\sigma)\overline{\mathbf{S}}\herm{\vphantom{\mathbf{S}}}_{X}(\sigma)\bigr\rvert\der\sigma\\
			&=\exp\int_{\mathrlap{0}}^{\mathrlap{\sfrac{1}{P}}} \ln\bigl\lvert\overline{\mathbf{C}}_{X}(\sigma)\bigr\rvert\der\sigma.
		\end{aligned}
	\end{equation}

\subsection{High SNR regime}

	If $\{x(n)\}$ is obtained from the additive model~\eqref{eq:additive}, we may relate the prediction and filtering \acrshort{mmse} as follows:
	\begin{align}
		&\mmse_{\mathrm{p}}(\snr)=\exp\int_{\mathrlap{0}}^{\mathrlap{\sfrac{1}{P}}} \ln\bigl\lvert\cycmat{DD}(\sigma)+\tfrac{1}{\snr}\id{P}\bigr\rvert\der\sigma\label{eq:causal_prediction}\\
		&\quad=\frac{\exp\int_0^{\sfrac{1}{P}}\ln\lvert\snr\cdot\cycmat{DD}(\sigma)+\id{P}\rvert\der\sigma}{\snr}=\frac{\euler^{\mmse_{\mathrm{c}}\cdot\snr}}{\snr}.\nonumber
	\end{align}
	With it, we can easily replicate the high \acrshort{snr} analysis from Section~\ref{ssec:sync_gain:high_snr} for prediction \acrshort{mmse}:	
	\begin{align}
		\smashoperator{\lim_{\snr\to\infty}} \mmse_{\mathrm{p}}(\snr) &= \smashoperator[l]{\lim_{\snr\to\infty}} \frac{\euler^{B\cdot\ln\snr}}{\snr} \cdot \euler^{\int_{0}^{B}\ln\klcyc_{\mathcal{D}}(\lambda)\der\lambda} \nonumber \\
		&\propto \frac{\snr^B}{\snr} = \frac{1}{\snr^{(1-B)}}.\label{eq:high_snr_pred}
	\end{align}
	Some remarks about this result should be made.
	The decay of~\eqref{eq:high_snr_pred} is slower than hyperbolic and is controlled by the occupied spectral band $B$.
	When $B\to0$, \ie very low $\omega_x^2$, it approaches the rate of decay of $\mmse_{\mathrm{nc}}$ at high \acrshort{snr}, as seen in~\eqref{eq:high_snr_non-causal}.
	When $B=1$, \ie $\{d(n)\}$ occupies the full \acrshort{kl} spectrum (high flatness), the desired signal becomes unpredictable and $\mmse_{\mathrm{p}}$ does not vanish at $\snr\to\infty$.

	\section{Numerical illustration}\label{sec:numerical}
		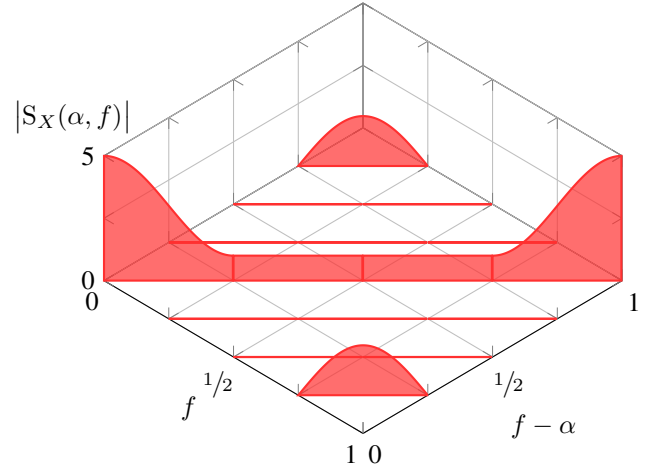
\begin{figure}[t]
	\centering
	\begin{tikzpicture}
	\pgfmathsetmacro{\xmax}{4}
	\pgfmathsetmacro{\ofset}{0}
	\pgfplotstableread[col sep=comma, row sep=newline]{Graphics/pam_psd.csv}{\taula}
	\begin{axis}[
		xmin=0-\ofset,	xmax=1+\ofset,
		ymin=0-\ofset,	ymax=1+\ofset,
		zmin=0,	zmax=5,
		view={45}{60},
		area plot/.style={
			fill opacity=0.7,
			draw=Firebrick1,
			fill=Firebrick1,
			mark=none,
			samples=200,
			samples y=1,
			smooth,
			domain=0:1/\xmax,
			thick
		},
		xlabel={$f$},
		ylabel={$f-\alpha$},
		zlabel={$\bigl\lvert\cyc_{X}(\alpha,f)\bigr\rvert$},
		xtick={0,0.25,0.5,0.75,1},
		ytick={0,0.25,0.5,0.75,1},
		ztick={0,2.5,5},
		xticklabels={0,,$\sfrac{1}{2}$,,1},
		yticklabels={0,,$\sfrac{1}{2}$,,1},
		zticklabels={0,,5},
		grid=major,
		axis lines = box,
		every x tick label/.style={anchor=near xticklabel},
		every y tick label/.style={anchor=near yticklabel},
		every z tick label/.style={anchor=near zticklabel},
		every axis x label/.style={
			at={(ticklabel cs:0.5)},
			anchor=near xticklabel,
			yshift=1em
		},
		every axis y label/.style={
			at={(ticklabel cs:0.5)},
			anchor=near yticklabel,
			yshift=0.75em
		},
		every axis z label/.style={
			at={(ticklabel cs:1)},
			anchor=south,
			yshift=0.5em,
		},
		clip=false,
		use units=false
		]
		\foreach \x in {1,2,3,4}
			\foreach \y in {1,2,3,4}
			{
				\addplot3[area plot] table [x=x\x_\y, y=y\x_\y, z=z\x_\y] {\taula} \closedcycle;
			}
	\end{axis}
\end{tikzpicture}
	\caption{Spectral correlation of $\{x(n)\}$ from model~\eqref{eq:model_num}, for $P=4$, $\Delta=0$ and $\power{z}=1$.}
	\label{fig:spec_corr}
\end{figure}

\noindent In this final section, we are going to apply the previous results to a linearly modulated digital communication system, in order to validate them numerically.
Consider a \acrfull{pam} signal of cycle period $P$, $\{d(n)\}$, affected by an independent additive noise component $\{z(n)\}$, which is \acrshort{wss} and white and has autocorrelation $\corr{z}(m)=\power{z}\delta_{m}$.
The transmitted complex baseband signal is constructed as follows:
\begin{equation}
	d(n)=\sum_{k\in\integs}a(k)\mathrm{b}(n-kP).
\end{equation}
The symbols $\{a(k)\}$ are uncorrelated and have power $P$ (\ie $\corr{a}(m)=P\delta_m$).
The shaping pulse $\mathrm{b}(n)$ is a 100\% excess bandwidth square-root raised cosine~\cite[Ex.~5-22]{Barry2004}, with \acrshort{dtft}
\begin{equation}
	\mathrm{B}(f)=\sqrt{P}\cos\bigl(\tfrac{\pi}{2}Pf\bigr),\quad f\in\bigl[-\tfrac{1}{P},\tfrac{1}{P}\bigr).
\end{equation}

The signal model of interest is as follows:
\begin{equation}
	x(n)=d(n-\varepsilon)+z(n),\label{eq:model_num}
\end{equation}
where $\varepsilon\sim\mathcal{U}[0,\Delta)$ is a random delay\footnote{A noninteger $\varepsilon$ represents a signal resampling in compact form~\cite[Sec.~11.1]{Proakis2013}.}.
The spectral correlation of $\{x(n)\}$ is derived in Appendix~\ref{app:spectral} and represented graphically in Fig.~\ref{fig:spec_corr}.
It is only defined on $\delta$-ridges parallel to the stationary manifold at $\alpha=\sfrac{k}{P}$ for $k=0,\dots,P-1$~\cite[Sec.~10.1.1]{Schreier2010}.
From~\eqref{eq:spec_corr_sims}, its cyclic spectrum is
\begin{equation}
	\cyc_{X}^{(\frac{k}{P})}(\sigma)=\power{z}\delta_k+\mathrm{B}(\sigma)\mathrm{B}^*\bigl(\sigma-\tfrac{k}{P}\bigr)\frac{1-\euler^{-\imunit2\pi\Delta\frac{k}{P}}}{\imunit2\pi\Delta\frac{k}{P}},\label{eq:cyc_spec_num}
\end{equation}
with which we construct the cyclic \acrshort{psd} matrix $\cycmat{X}(\sigma)$ for all $\sigma\in[0,\sfrac{1}{P})$, as explained in Section~\ref{sec:kl} (see Fig.~\ref{fig:psd}).
By obtaining its eigenvalues $\klcycmat{\mathcal{X}}(\sigma)$ for each $\sigma$, the \acrshort{kl}-\acrshort{psd} can be derived, since $[\klcycmat{\mathcal{X}}(\sigma)]_{p,p}=\klcyc_{\mathcal{X}}(\sigma+\sfrac{p}{P})$.

The maximum delay $\Delta$ allows to control the statistical properties of $\{x(n)\}$.
Notice that its impact cannot be observed on the \acrshort{psd}; indeed, if we take~\eqref{eq:cyc_spec_num} and set $k=0$, we obtain $\cyc_{X}(f)=\power{z}+\lvert\mathrm{B}(f)\rvert^2$, which does not depend on $\Delta$.
On the contrary, the \acrshort{kl}-\acrshort{psd} does behave differently for different values of $\Delta$, as shown in~Fig.~\ref{fig:psd_klpsd}.
In particular, Fig.~\ref{fig:kl-psd} shows $\klcyc_{\mathcal{X}}(\lambda)$ obtained directly from $\klcycmat{\mathcal{X}}(\sigma)$ (which contains the eigenvalues of $\cycmat{X}(\sigma)$ in decreasing order) as $\klcyc_{\mathcal{X}}(\sigma+\sfrac{p}{P})=[\klcycmat{\mathcal{X}}(\sigma)]_{p,p}$, whereas Fig.~\ref{fig:sorted} displays its \textit{decreasing rearrangement}~\cite[Def.~D.1.]{Marshall2011}.
For $\Delta=0$, (\ie $\{d(n)\}$ is fully \acrshort{cs}), the signal component of $\klcyc_{\mathcal{X}}(\lambda)$ is contained in a spectral band of width $\sfrac{1}{P}$, while the rest of the spectrum is only occupied by noise.
As $\Delta$ increases (\eg $\Delta=\sfrac{P}{6}$), some amount of the signal \acrshort{kl}-\acrshort{psd} leaks to the band $[\sfrac{1}{P},\sfrac{2}{P})$.
When $\Delta=\sfrac{P}{(P-1)}$, the leakage is maximum and $\{d(n)\}$ becomes \acrshort{wss}, in which case the \acrshort{kl}-\acrshort{psd} decreasing rearrangement coincides with the \acrshort{cl}-\acrshort{psd} one.
\begin{figure}[t]
	\centering
	\begin{subfloat}[\acrshort{cl}-\acrshort{psd} and \acrshort{kl}-\acrshort{psd} for various values of $\Delta$ (see legend in Fig.~\ref{fig:sorted}).]
	{
		\centering
		\begin{tikzpicture}
	\pgfplotstableread[col sep=comma, row sep=newline]{Graphics/pam_spectra.csv}{\taula}
	\begin{axis}[
		width=\linewidth,	height=.5\linewidth,
		xmin=0,				ymin=0,
		xmax=1,				ymax=5.5,
		xlabel=$\lambda$,		ylabel=\acrshort{psd},
		xtick={0,0.25,...,1},	ytick={0,1,...,5},
		xticklabels={$0$,$\sfrac{1}{P}$,$\sfrac{2}{P}$,$\sfrac{3}{P}$,$1$},
		axis on top,
		axis line style={thick},
		linestyle/.style={#1, very thick, mark size=2pt, mark repeat=50},
		areastyle/.style={#1},
		every axis plot/.style={/pgfplots/on layer=axis foreground},
		axis background/.style={fill=white},
		legend cell align=left,
		legend columns=1,
		legend transposed,
		legend style={at={(0.83,0.98)}},
		x label style={at={(axis description cs:0.5,-0.15)},anchor=north},
		use units=false
		]
		\addplot[name path=kl0, linestyle=DeepSkyBlue3, mark=+] table [y=KL0] {\taula};
		\addplot[name path=kl1, linestyle=SpringGreen3, mark=o] table [y=KL1] {\taula};
		\addplot[name path=kl2, linestyle=Firebrick3, mark=x] table [y=KL2] {\taula};
		\addplot[name path=cl, linestyle=Gold3, dashed] table [y=CL] {\taula};
		\addplot[name path=zero, line width=0, samples=2, domain=0:1] {0};
		\addplot[areastyle=DeepSkyBlue1!50!white] fill between [of=kl0 and kl1, soft clip={domain=0:0.245}];
		\addplot[areastyle=SpringGreen1!50!white] fill between [of=kl1 and kl2, soft clip={domain=0:0.245}];
		\addplot[areastyle=Firebrick1!50!white] fill between [of=kl2 and cl, soft clip={domain=0.125:0.25}];
		\addplot[areastyle=Firebrick1!50!white] fill between [of=kl2 and kl0, soft clip={domain=0.25:0.5}];
		\addplot[areastyle=Gold1!50!white] fill between [of=cl and zero];
	\end{axis}
\end{tikzpicture}
		\label{fig:kl-psd}
	}
	\end{subfloat}
	\begin{subfloat}[Decreasing rearrangement of densities from Fig.~\ref{fig:kl-psd}.]
	{
		\centering
		\begin{tikzpicture}
	\pgfplotstableread[col sep=comma, row sep=newline]{Graphics/pam_spectra.csv}{\taula}
	\begin{axis}[
		width=\linewidth,	height=.5\linewidth,
		xmin=0,				ymin=0,
		xmax=1,				ymax=5.5,
		xlabel=$\lambda$,		ylabel=\acrshort{psd},
		xtick={0,0.25,...,1},	ytick={0,1,...,5},
		xticklabels={$0$,$\sfrac{1}{P}$,$\sfrac{2}{P}$,$\sfrac{3}{P}$,$1$},
		axis on top,
		axis line style={thick},
		linestyle/.style={#1, very thick, mark size=2pt, mark repeat=50},
		areastyle/.style={#1},
		every axis plot/.style={/pgfplots/on layer=main},
		axis background/.style={fill=white},
		legend cell align=left,
		legend columns=1,
		legend transposed,
		x label style={at={(axis description cs:0.5,-0.15)},anchor=north},
		use units=false
		]
		\addplot[name path=kl0, linestyle=DeepSkyBlue3, mark=+] table [y=KL0_sort] {\taula};
		\addlegendentry{$\Delta=0$};
		\addplot[name path=kl1, linestyle=SpringGreen3, mark=o] table [y=KL1_sort] {\taula};
		\addlegendentry{$\Delta=\sfrac{P}{6}$};
		\addplot[name path=kl2, linestyle=Firebrick3, mark=x] table [y=KL2_sort] {\taula};
		\addlegendentry{$\Delta=\sfrac{P}{(P-1)}$};
		\addplot[name path=cl, linestyle=Gold3, dashed] table [y=CL_sort] {\taula};
		\addlegendentry{\acrshort{cl}-\acrshort{psd}};
		\addplot[name path=zero, line width=0, samples=2, domain=0:1] {0};
		\addplot[areastyle=DeepSkyBlue1!50!white] fill between [of=kl0 and kl1, soft clip={domain=0:0.25}];
		\addplot[areastyle=SpringGreen1!50!white] fill between [of=kl1 and kl2, soft clip={domain=0:0.25}];
		\addplot[areastyle=Gold1!50!white] fill between [of=cl and zero];
	\end{axis}
\end{tikzpicture}
		\label{fig:sorted}
	}
	\end{subfloat}
	\caption{Spectral densities of $\{x(n)\}$ for $P=4$ and $\power{z}=1$.}
	\label{fig:psd_klpsd}
\end{figure}
\begin{figure}[t]
	\centering
	\begin{tikzpicture}
	\pgfplotstableread[col sep=comma, row sep=newline]{Graphics/rep_entro.csv}{\taula}
	\begin{axis}[
		width=.9\linewidth,	height=.35\linewidth,
		xmin=0,				ymin=-0.35,
		xmax=4/3,			ymax=-0.2,
		xlabel=$\Delta$,			ylabel=$\mathrm{h}(\{\phi\})$,
		xtick={0,1/3,...,4/3},
		xticklabels={$0$,$\sfrac{1}{3}$,$\sfrac{2}{3}$,$1$,$\frac{P}{P-1}$},
		axis on top,
		axis line style={thick},
		linestyle/.style={#1, very thick, mark size=2pt, mark repeat=20},
		every axis plot/.style={/pgfplots/on layer=main},
		axis background/.style={fill=white},
		legend cell align=left,
		legend pos=south east,
		y unit=nat,
		x label style={at={(axis description cs:0.5,-0.25)},anchor=north}
		]
		\addplot[linestyle=Firebrick2, dashed] table [y=CL] {\taula};
		\addlegendentry{\acrshort{cl}};
		\addplot[linestyle=DeepSkyBlue2, mark=o] table [y=rep_entro] {\taula};
		\addlegendentry{\acrshort{kl}};
	\end{axis}
\end{tikzpicture}
	\vspace{-1em}
	\caption{Representation entropy of $\{x(n)\}$.}
	\label{fig:rep_entropy}
\end{figure}
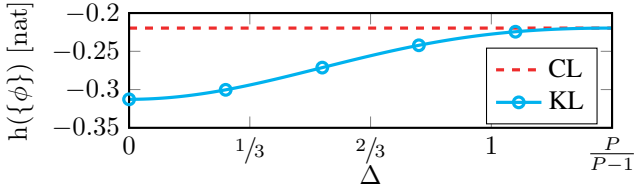

A different way to analyze the effect of $\Delta$ onto the statistical behavior of $\{x(n)\}$ is through the representation entropy from Section~\ref{ssec:optimal} (Fig.~\ref{fig:rep_entropy}). 
As expected, the lowest entropy is achieved for $\Delta=0$, since the \acrshort{kl}-\acrshort{psd} of $\{x(n)\}$ is the most compact (\ie \acrshort{cs}).
Its value increases with $\Delta$ until it reaches a maximum for $\Delta=\sfrac{P}{(P-1)}$, which coincides with the entropy of the \acrshort{cl} representation of $\{x(n)\}$, as it is \acrshort{wss}.

\subsection{Signal processing performance}
	
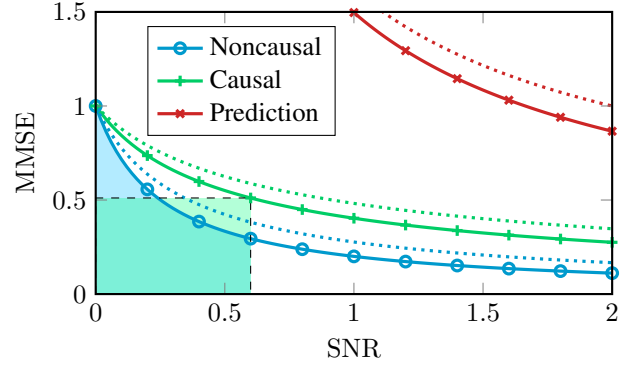
\begin{figure}[t]
		\centering
		\begin{tikzpicture}
	\pgfplotstableread[col sep=comma, row sep=newline]{Graphics/wiener.csv}{\taula}
	\begin{axis}[
		width=.95\linewidth,	height=.6\linewidth,
		xmin=0,				ymin=0,
		xmax=2,				ymax=1.5,
		xlabel=$\snr$,		ylabel=$\mmse$,
		xtick={0,0.5,...,2},	ytick={0,0.5,1,1.5},
		axis on top,
		axis line style={thick},
		linestyle/.style={#1, very thick, mark size=2pt, mark repeat=10},
		every mark/.append style={solid},
		axis background/.style={fill=white},
		legend cell align=left,
		every axis legend/.append style={at={(0.1,0.95)}, anchor=north west},
		use units=false
		]
		\fill[opacity=0.3, color=SpringGreen1] (0,0) rectangle (0.6,0.5109262427022334);
		\addplot[linestyle=DeepSkyBlue3, mark=o, name path=nc] table [y=MMSE_nc] {\taula};
		\addlegendentry{Noncausal};
		\addplot[linestyle=SpringGreen3, mark=+] table [y=MMSE_c] {\taula};
		\addlegendentry{Causal};
		\addplot[linestyle=Firebrick3, mark=x] table [y=MMSE_p] {\taula};
		\addlegendentry{Prediction};
		\addplot[linestyle=DeepSkyBlue3, dotted] table [y=MMSE_nc_WSS] {\taula};
		\addplot[linestyle=SpringGreen3, dotted] table [y=MMSE_c_WSS] {\taula};
		\addplot[linestyle=Firebrick3, dotted] table [y=MMSE_p_WSS] {\taula};
		\addplot[name path=zero, line width=0, samples=2, domain=0:0.6] {0};
		\addplot[DeepSkyBlue1, opacity=0.3] fill between [of=nc and zero, soft clip={domain=0:0.6}];
		\draw[draw=black, dashed] (0,0) rectangle (0.6,0.5109262427022334);
	\end{axis}
\end{tikzpicture}
		\vspace{-1em}
		\caption{\acrshort{mmse} in synchronous (solid lines) and asynchronous (dotted lines) processing. $\Delta=0$ and $P=4$.}
		\label{fig:MSE_gsv}
	\end{figure}

	We are now interested in numerically assessing the performance of smoothing, filtering and prediction, given model~\eqref{eq:model_num}.
	Fig.~\ref{fig:MSE_gsv} displays the \acrshort{mmse} of both synchronous and asynchronous processing in the three modalities for different \acrshort{snr} values, making very clear the gains achievable by exploiting cyclostationarity.
	The relationship between non-causal and causal filtering from~\cite[Th.~8]{Guo2005}, explored in Section~\ref{sec:causal}, is illustrated as well.
	The area under $\mmse_{\text{nc}}$ in the interval $\snr\in[0,0.6]$ is the same as the one in the colored box, which coincides with $\mmse_{\text{c}}$ at $\snr=0.6$, as expected from~\eqref{eq:causal_mmse}.
	
\begin{figure}[t]
		\centering
		\begin{tikzpicture}
	\pgfplotstableread[col sep=comma, row sep=newline]{Graphics/wiener_high.csv}{\taula}
	\begin{semilogyaxis}[
		width=.95\linewidth,	height=.8\linewidth,
		xmin=20,			ymin=1e-1,
		xmax=50,			ymax=1e3,
		xlabel=$\snr$,		ylabel=$\mse\times\snr$,
		xtick={20,30,...,50},
		axis on top,
		axis line style={thick},
		linestyle/.style={#1, very thick, forget plot},
		markstyle/.style={#1, mark size=2pt, mark repeat=10, only marks, very thick},
		every mark/.append style={solid},
		axis background/.style={fill=white},
		legend cell align=left,
		legend pos=north west,
		legend columns=3,
		transpose legend,
		x unit=dB
		]
		\addplot[linestyle=DeepSkyBlue3] table [y=MMSE_nc_2] {\taula};
		\addplot[linestyle=SpringGreen3] table [y=MMSE_c_2] {\taula};
		\addplot[linestyle=Firebrick3] table [y=MMSE_p_2] {\taula};
		\addplot[linestyle=DeepSkyBlue3, dashed] table [y=MMSE_nc_4] {\taula};
		\addplot[linestyle=SpringGreen3, dashed] table [y=MMSE_c_4] {\taula};
		\addplot[linestyle=Firebrick3, dashed] table [y=MMSE_p_4] {\taula};
		\addplot[markstyle=DeepSkyBlue3, mark=o] table [y=MMSE_nc_2] {\taula};
		\addplot[markstyle=SpringGreen3, mark=+] table [y=MMSE_c_2] {\taula};
		\addplot[markstyle=Firebrick3, mark=x] table [y=MMSE_p_2] {\taula};
		\addplot[markstyle=DeepSkyBlue3, mark=o, forget plot] table [y=MMSE_nc_4] {\taula};
		\addplot[markstyle=SpringGreen3, mark=+, forget plot] table [y=MMSE_c_4] {\taula};
		\addplot[markstyle=Firebrick3, mark=x, forget plot] table [y=MMSE_p_4] {\taula};
		\legend{Noncausal,Causal,Prediction};
		\addplot[DeepSkyBlue3, samples=2, forget plot, domain=20:50] {1/(2*(10^(x/10)))*10^(x/10)};
		\addplot[DeepSkyBlue3, samples=2, forget plot, domain=20:50, dashed] {1/(4*(10^(x/10)))*10^(x/10)};
		\addplot[SpringGreen3, samples=101, forget plot, domain=20:50] {(1/2)*x*ln(10)/(10^(1+x/10))*10^(x/10)};
		\addplot[SpringGreen3, samples=101, forget plot, domain=20:50, dashed] {(1/4)*x*ln(10)/(10^(1+x/10))*10^(x/10)};
		\addplot[Firebrick3, samples=101, forget plot, domain=20:50] {10^(-x*(1-1/2)/10)*10^(x/10)};
		\addplot[Firebrick3, samples=101, forget plot, domain=20:50, dashed] {10^(-x*(1-1/4)/10)*10^(x/10)};
		\addlegendimage{black, line legend, very thick};
		\label{k2};
		\addlegendimage{black, line legend, very thick, dashed};
		\label{k4};
		\node[draw, fill=white, anchor=north west] at (rel axis cs: 0.35,0.95) {\shortstack[l]{\ref*{k2} $P=2$ \\ \ref*{k4} $P=4$}};
		\node[color=Firebrick3] at (35,25) {\footnotesize slope: $\sfrac{1}{P}$};
		\node[color=SpringGreen3] at (35,1) {\footnotesize slope: $O(\sfrac{1}{\text{\acrshort{snr}}[\unit{\dB}]})$};
		\node[color=DeepSkyBlue3] at (35,0.15) {\footnotesize slope: \num{0}};
	\end{semilogyaxis}
\end{tikzpicture}
		\vspace{-1em}
		\caption{\acrshort{mmse}$\times$\acrshort{snr} vs. \acrshort{snr} for synchronous processing. The dotted lines are the corresponding approximations in the high \acrshort{snr} regime.}
		\label{fig:MSE_high}
	\end{figure}
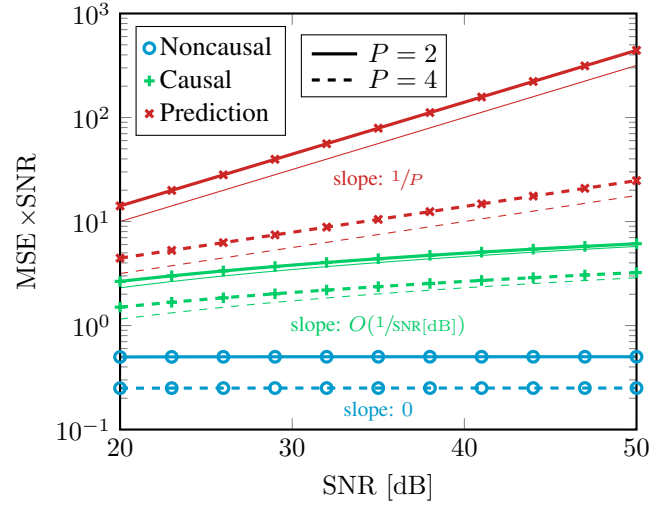

	In Fig.~\ref{fig:MSE_high}, there is a plot of \acrshort{mmse}\texttimes\acrshort{snr} against \acrshort{snr} for synchronous processing in the three modalities.
	Regarding smoothing, its corresponding curve is a horizontal straight line, since its \acrshort{mmse} decays hyperbolically.
	We can observe the high \acrshort{snr} approximation derived in Section~\ref{ssec:sync_gain:high_snr} is remarkably accurate.
	Its ordinate is dictated by the bandwidth parameter in~\eqref{eq:high_snr_non-causal} which, in the model considered, depends on the signal period (\ie $B=\sfrac{1}{P}$).
	Relating to causal filtering, its curve increases at a slower rate the higher the \acrshort{snr} is.
	This implies its \acrshort{mmse} presents asymptotically hyperbolic decay, as predicted in~\eqref{eq:high_snr_causal}.
	Its approximation improves as the \acrshort{snr} grows, and the $O\bigl(\snr^{-1}\bigr)$ term from~\eqref{eq:high_snr_causal} vanishes. 
	Finally, the prediction curves increase linearly with \acrshort{snr}, with a slope equal to $\sfrac{1}{P}$: this is the performance loss with respect to hyperbolic decay, as expected from~\eqref{eq:high_snr_pred}.
	Notice that the approximation curve displays the same slope as the real one.
	
	
\begin{figure}[t]
		\centering
		\begin{tikzpicture}
	\pgfplotstableread[col sep=comma, row sep=newline]{Graphics/sync_gain.csv}{\taula}
	\pgfplotsset{set layers}
	\begin{semilogyaxis}[
		width=\linewidth,	height=.5\linewidth,
		xmin=0,			ymin=1,
		xmax=1,			ymax=16,
		xlabel=$\Delta\times\sfrac{(P-1)}{P}$,		ylabel=$1/\zeta$,
		xtick={0,0.25,...,1},
		ytick={1, 2, 4, 8, 16},
		axis on top,
		axis line style={thick},
		linestyle/.style={#1, very thick, forget plot},
		markstyle/.style={#1, mark size=2pt, mark repeat=20, only marks, very thick},
		every mark/.append style={solid},
		axis background/.style={fill=white},
		legend cell align=left,
		legend pos=north east,
		legend columns=3,
		transpose legend,
		log basis y=2,
		mark layer=like plot,
		use units=false
		]
		\addplot[linestyle=DeepSkyBlue3] table [y=MMSE_nc_2] {\taula};
		\addplot[linestyle=DeepSkyBlue3, dashed] table [y=MMSE_nc_4] {\taula};
		\addplot[linestyle=DeepSkyBlue3, dotted] table [y=MMSE_nc_8] {\taula};
		\addplot[markstyle=DeepSkyBlue3, mark=o] table [y=MMSE_nc_2] {\taula};
		\addplot[markstyle=DeepSkyBlue3, mark=o, forget plot] table [y=MMSE_nc_4] {\taula};
		\addplot[markstyle=DeepSkyBlue3, mark=o, forget plot] table [y=MMSE_nc_8] {\taula};	
		
		\addplot[linestyle=SpringGreen3] table [y=MMSE_c_2] {\taula};
		\addplot[linestyle=SpringGreen3, dashed] table [y=MMSE_c_4] {\taula};
		\addplot[linestyle=SpringGreen3, dotted] table [y=MMSE_c_8] {\taula};
		\addplot[markstyle=SpringGreen3, mark=+] table [y=MMSE_c_2] {\taula};
		\addplot[markstyle=SpringGreen3, mark=+, forget plot] table [y=MMSE_c_4] {\taula};
		\addplot[markstyle=SpringGreen3, mark=+, forget plot] table [y=MMSE_c_8] {\taula};
		
		\addplot[linestyle=Firebrick3] table [y=MMSE_p_2] {\taula};		
		\addplot[linestyle=Firebrick3, dashed] table [y=MMSE_p_4] {\taula};
		\addplot[linestyle=Firebrick3, dotted] table [y=MMSE_p_8] {\taula};
		\addplot[markstyle=Firebrick3, mark=x] table [y=MMSE_p_2] {\taula};
		\addplot[markstyle=Firebrick3, mark=x, forget plot] table [y=MMSE_p_4] {\taula};
		\addplot[markstyle=Firebrick3, mark=x, forget plot] table [y=MMSE_p_8] {\taula};
		\legend{Noncausal,Causal,Prediction};
		\addlegendimage{black, line legend, very thick};
		\label{sgain_k2};
		\addlegendimage{black, line legend, very thick, dashed};
		\label{sgain_k4};
		\addlegendimage{black, line legend, very thick, dotted};
		\label{sgain_k8};
		\node[draw, fill=white, anchor=north east] at (rel axis cs: 0.65,0.95) {\shortstack[l]{\ref*{sgain_k2} $P=2$ \\ \ref*{sgain_k4} $P=4$ \\ \ref*{sgain_k8} $P=8$}};
	\end{semilogyaxis}
\end{tikzpicture}
		\vspace{-1em}
		\caption{Processing gain in terms of synchronization reliability at $\snr=30\text{dB}$.}
		\label{fig:sync_gain}
	\end{figure}
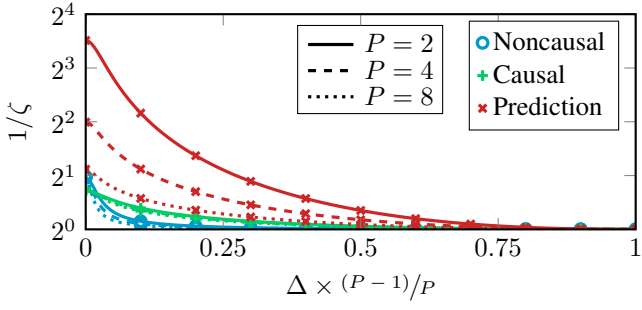

	To assess the gain achievable by performing a synchronous treatment of the signal at high \acrshort{snr} we refer to~Fig.~\ref{fig:sync_gain}.
	The horizontal axis is a scan across all the possible values of $\Delta\in[0,\sfrac{P}{(P-1)}]$; it represents the synchronization reliability at the receiver (\ie 0 is perfect synchronization while 1 is maximum uncertainty).
	In both smoothing and filtering, the achievable gain is not affected by the signal period $P$ when perfect synchronization is available.
	The gain degradation as the uncertainty increases, however, is less sensitive to $\Delta$ in filtering.
	The prediction case bears some comment.
	Under perfect synchronization, the achievable gains are the highest among the three processing modes and are remarkably affected by $P$: the higher the period, the lower the gain.
	This phenomenon can be explained in terms of spectral flatness.
	For $P=2$, the \acrshort{kl}-\acrshort{psd} of $\{d(n)\}$ is concentrated in a band of width $1/2$.
	On the contrary, its \acrshort{cl}-\acrshort{psd} occupies the full spectrum, \ie it is significantly flatter.
	Therefore, implementing a perfectly synchronized processing yields a notable performance gain.
	As $P$ increases, however, the \acrshort{cl}-\acrshort{psd} concentrates in a smaller band and the synchronous gain shrinks.
	
	\section{Conclusions}\label{sec:conclusion}
		This work has explored the problem of synchronous processing of \acrshort{cs} signals.
By using their \acrshort{kl} representation, we have leveraged its desirable properties to conduct various theoretical analyses. In particular, its uncorrelated eigendecomposition and time-shift invariant spectral density have proven to be very valuable in the extraction of asymptotic performance bounds for filtering applications.

This spectral treatment has enabled to address smoothing, filtering and prediction under the same unified framework and study cyclic Wiener processing without having to delve into specific architectures.
The unitary transformation that connects the \acrshort{cl} and \acrshort{kl} expansions has played a central role in obtaining compact \acrshort{mmse} expressions, as integrals of traces and determinants of the cyclic spectrum matrix.
With them, we have also quantified the achievable synchronous gain, which is of remarkable interest in prediction applications.
Indeed, the optimal energy compaction of the \acrshort{kl} spectrum is intimately linked to improved predictability.

The toolset developed in the present work can be applied in various new directions.
Interesting extensions can be found in \textit{almost \acrshort{cs}} processes~\cite{Napolitano2013,Napolitano2016a}, which are essential in state-of-the-art studies of signal periodicities. 
From a theoretical point of view, the presence of coherence statistics in \acrshort{mmse} expressions should be investigated further.
This would provide deep insights in the relation between synchronous signal processing and other problems in information theory.

	\appendix
		\subsection{Proof of Theorem~\ref{thm:cs}}\label{app:cs}

    The following proof is an alternative version of the one presented in~\cite{Riba2022}.
    It is based on checking that~\eqref{eq:eigenequation} holds for \acrshort{cs} processes with the proposed basis:
    \begin{equation}
	   	\klcyc_{\mathcal{X}}^{(p)}(\sigma) \phi_{\mathrm{CS}}^{(p)}(k,\sigma) \der\sigma = \smashoperator[l]{\lim_{N\to\infty}} \frac{1}{N} \smashoperator{\sum_{l=-\frac{N}{2}}^{\frac{N}{2}-1}} \corr{x}(l,k-l) \phi_{\mathrm{CS}}^{(p)}(l,\sigma).
    \end{equation}
    Applying~\eqref{eq:autocorr} and the \acrshort{cl} expansion of $\{x(k)\}$, we have
    \begin{align}
		&\klcyc_{\mathcal{X}}^{(p)}(\sigma) \phi_{\mathrm{CS}}^{(p)}(k,\sigma) \der\sigma\\
		&\quad= \smashoperator[l]{\lim_{N\to\infty}} \frac{1}{N} \sum_{l} \expec[x(k)x^*(l)] \smashoperator{\sum_{c=0}^{P-1}} b_{X}^{(p)}(c,\sigma) \euler^{\imunit2\pi(\sigma+\frac{c}{P})l} \nonumber\\
		&\quad= \sum_{c} b_{X}^{(p)} (c,\sigma) \expec\bigg[\bigg(\sum_{r=0}^{P-1} \int_{\mathrlap{0}}^{\mathrlap{\sfrac{1}{P}}} \euler^{\imunit2\pi(\sigma'+\frac{r}{P})k} \der\nu_x^{(r)}(\sigma')\bigg)\nonumber\\
		&\qquad\cdot \bigg(\lim_{N\to\infty} \frac{1}{N} \sum_{l} x(l) \euler^{-\imunit2\pi l(\sigma+\frac{c}{P})}\bigg)^*\bigg] \nonumber\\
		&\quad= \sum_{r,c} b_{X}^{(p)}(c,\sigma) \int_{\mathrlap{0}}^{\mathrlap{\sfrac{1}{P}}} \euler^{\imunit2\pi(\sigma'+\frac{r}{P})k} \expec\bigl[\der\nu_x^{(r)}(\sigma')\der\nu_x^{(c)*}(\sigma)\bigr], \nonumber
    \end{align}
    where $\der\nu^{(p)}_x(\sigma)$ is defined in~\eqref{eq:cyc_psd_mat}.
    Using the spectral correlation of a \acrshort{cs} process~\eqref{eq:cs_spec_corr} yields:
   	\begin{equation}
   		\begin{aligned}
   			&\klcyc_{\mathcal{X}}^{(p)}(\sigma) \phi_{\mathrm{CS}}^{(p)}(k,\sigma) \der\sigma = \sum_{r,c} b_{X}^{(p)}(c,\sigma) \der\sigma\\
   			&\qquad\cdot \int_{\mathrlap{0}}^{\mathrlap{\sfrac{1}{P}}} \euler^{\imunit2\pi(\sigma'+\frac{r}{P})k} \cyc_{X}^{(\frac{r-c}{P})}(\sigma'+\tfrac{r}{P}) \delta(\sigma'-\sigma) \der\sigma'\\
    		&\quad= \sum_{r,c} \euler^{\imunit2\pi(\sigma+\frac{r}{P})k} b_{X}^{(p)}(c,\sigma) \cyc_{X}^{(\frac{r-c}{P})}(\sigma+\tfrac{r}{P}) \der\sigma.
	   	\end{aligned}
   	\end{equation}
    Afterwards, we substitute the proposed basis on the LHS of the equality and cancel the common exponential terms that depend on $\sigma$:
    \begin{multline}
		\klcyc_{\mathcal{X}}^{(p)}(\sigma) \sum_{r} b_{X}^{(p)}(r,\sigma)\euler^{\imunit2\pi\frac{r}{P}k}\\
		= \sum_{r,c} \euler^{\imunit2\pi\frac{r}{P}k}b_{X}^{(p)}(c,\sigma) \cyc_{X}^{(\frac{r-c}{P})}(\sigma+\tfrac{r}{P}).
    \end{multline}
    Finally, each term of the sum with respect to $r$ is equated and stacked in vector form:
    \begin{equation}
	    \begin{aligned}
		    \klcyc_{\mathcal{X}}^{(p)}(\sigma) b_{X}^{(p)}(r,\sigma) &= \sum_{c} b_{X}^{(p)}(c,\sigma) \cyc_{X}^{(\frac{r-c}{P})}(\sigma+\tfrac{r}{P})\\
		    \klcyc_{\mathcal{X}}^{(p)}(\sigma) \mathbf{b}_{X}^{(p)}(\sigma) &= \cycmat{X}(\sigma) \mathbf{b}_{X}^{(p)}(\sigma).
	    \end{aligned}
    \end{equation}
    In order to fulfill this eigenequation, $\mathbf{b}_X^{(p)}(\sigma)$ must be an eigenvector of $\cycmat{X}(\sigma)$ and $\klcyc_{\mathcal{X}}^{(p)}(\sigma)$ its corresponding eigenvalue.
    This completes the proof.
    \hfill\IEEEQED

\subsection{Proof of Corollary~\ref{cor:cl-kl1}}\label{app:cl-kl}
    The proof is obtained by particularizing~\eqref{eq:kl} for the \acrshort{cs} basis~\eqref{eq:cs_basis} and applying the definition of the \acrshort{cl} transform~\eqref{eq:icl}:
    \begin{align}
		&\der\xi_x^{(p)}(\sigma) = \smashoperator[l]{\lim_{N\to\infty}} \frac{1}{N} \smashoperator{\sum_{n=-\frac{N}{2}}^{\frac{N}{2}-1}} x(n) \smashoperator{\sum_{q=0}^{P-1}} b_X^{(p)*}(q,\sigma) \euler^{-\imunit2\pi(\sigma+\frac{q}{P})n} \nonumber\\
		&\quad= \sum_{q} b_X^{(p)*}(q,\sigma) \Bigl(\lim_{N\to\infty} \frac{1}{N} \sum_{n} x(n) \euler^{-\imunit2\pi(\sigma+\frac{q}{P})n} \Bigr) \nonumber\\
		&\quad= \sum_{q} b_X^{(p)*}(q,\sigma) \der\nu^{(q)}_x(\sigma) = \mathbf{b}_X^{(p)\mathrm{H}}(\sigma) \Breve{\mathbf{x}}(\sigma). \label{eq:cl+decorr}
    \end{align}
    \hfill\IEEEQED

\subsection{Time-shift effect on the KL transform of CS processes}\label{app:time-shift}
    Changing the reference time of a \acrshort{cs} process adds a phase shift to its \acrshort{kl} representation:
    \begin{align}
     	&\der\xi_{x,\mathrm{CS}}^{(p)}(\sigma,n_0) = \smashoperator[l]{\lim_{N\to\infty}} \frac{1}{N} \smashoperator{\sum_{n=-\frac{N}{2}}^{\frac{N}{2}-1}} x(n+n_0) \smashoperator{\sum_{q=0}^{P-1}} b_{X}^{(p)*}(q,\sigma) \nonumber\\
     	&\qquad\cdot \euler^{-\imunit2\pi\frac{q}{P}n_0} \euler^{-\imunit2\pi(\sigma+\frac{q}{P})n} \nonumber\\
       	&\quad= \smashoperator[l]{\lim_{N\to\infty}} \tfrac{1}{N} \cdot \smashoperator{\sum_{n'=-\frac{N}{2}+n_0}^{\frac{N}{2}-1+n_0}} x(n') \sum_{q} b_X^{(p)*}(q,\sigma) \euler^{-\imunit2\pi(\sigma+\frac{q}{P})n'} \euler^{\imunit2\pi\sigma n_0} \nonumber\\
       	&\quad= \der\xi_{x,\mathrm{CS}}^{(p)}(\sigma) \euler^{\imunit2\pi\sigma n_0}.
    \end{align}
    
\subsection{Proof of Theorem~\ref{thm:pred}}\label{app:pred}
	
	This proof is a refinement on the one provided in~\cite{Riba2022}, adapted to the present notation and setting.
	Recall~\eqref{eq:determinants}:
	\begin{equation}
		\bigl\lvert\corrmat_{x}^{(N+1)}(n)\bigr\rvert = \mmse_{\mathrm{p}}^{(N)}(n) \cdot \bigl\lvert\corrmat_{x}^{(N)}(n-1)\bigr\rvert.
	\end{equation}
	Developing $\bigl\lvert\corrmat_{x}^{(N)}(n-1)\bigr\rvert$ allows to express it as a recursion~\cite[Sec.~2.4.3]{Vaidyanathan2007}:
	\begin{equation}\label{eq:recursion}
		\bigl\lvert\corrmat_{x}^{(N+1)}(n)\bigr\rvert = \smashoperator{\prod_{l=0}^{N}} \mmse_{\mathrm{p}}^{(l)}(n-N+l).
	\end{equation}
	
	Since $\{\corrmat_{x}\}$ are \textit{$P$-Toeplitz}~\cite{Riba2022}, it is known that~\cite{Gyires1962,Widom1974}
	\begin{equation}\label{eq:widom}
		\lim_{N\to\infty} \frac{\bigl\lvert\corrmat_{x}^{(N+1)}(n)\bigr\rvert}{\bigl\lvert\corrmat_{x}^{(N+1-P)}(n-P)\bigr\rvert} = \exp\int_{\mathrlap{0}}^{\mathrlap{1}} \ln\bigl\lvert\mathbf{T}_{x}(\lambda)\bigr\rvert\der\lambda,
	\end{equation}
	where $\mathbf{T}_{x}(\lambda)\in\complex^{P\times P}$ is a matrix that contains the \textit{Rihaczek spectrum}~\cite[Sec.~1.4.1.3]{Hlawatsch2011} of $\{x(n)\}$.
	It can obtained as the \acrshort{dtft} of the sequence of $P\times P$ blocks of $\{\corrmat_x\}$~\cite{VilaInsa2025b}.
	Using the recursion formula~\eqref{eq:recursion}, the LHS of~\eqref{eq:widom} reduces to
	\begin{align}
		&\lim_{N\to\infty} \frac{\prod_{l=0}^N \mmse_{\mathrm{p}}^{(l)}(n-N+l)}{\prod_{l'=0}^{N-P} \mmse_{\mathrm{p}}^{(l')}(n-N+l')} \\
		&\qquad= \lim_{N\to\infty} \smashoperator[r]{\prod_{l=1}^{P}} \mmse_{\mathrm{p}}^{(l+N-P)}(n+l-P) = \mmse_{\mathrm{p}}^P. \nonumber
	\end{align}
	Therefore, the asymptotic \acrshort{mse} is
	\begin{equation}
		\mmse_{\mathrm{p}} = \exp\Bigl(\frac{1}{P}\int_{\mathrlap{0}}^{\mathrlap{1}} \ln\bigl\lvert\mathbf{T}_{x}(\lambda)\bigr\rvert\der\lambda\Bigr).
	\end{equation}
	Since the sequence $\{\lvert\corrmat_{x}^{(N+1)}(n)\rvert/\lvert\corrmat_{x}^{(N+1-P)}(n-P)\rvert\}$ is monotonically nonincreasing with $N$~\cite{Gyires1962}, this is a lower bound on the achievable \acrshort{mmse} of one-step prediction of a \acrshort{cs} signal.
	
	The Rihaczek spectrum matrix is related to the cyclic \acrshort{psd} matrix as follows~\cite{VilaInsa2025b}:
	\begin{equation}
		\mathbf{T}_x(\lambda) = \mathbf{U}\bigl(\sfrac{\lambda}{P}\bigr)\cycmat{x}\bigl(\sfrac{\lambda}{P}\bigr)\herm{\mathbf{U}}\bigl(\sfrac{\lambda}{P}\bigr),
	\end{equation}
	where $\mathbf{U}(\lambda)\in\complex^{P\times P}$ are unitary matrices.
	Therefore, the prediction \acrshort{mmse} may be expressed as
	\begin{equation}
		\mmse_{\mathrm{p}} = \exp\Bigl(\frac{1}{P}\int_{\mathrlap{0}}^{\mathrlap{1}} \ln\bigl\lvert\cycmat{x}\bigl(\tfrac{\lambda}{P}\bigr)\bigr\rvert\der\lambda\Bigr) = \exp\int_{\mathrlap{0}}^{\mathrlap{\sfrac{1}{P}}} \ln\bigl\lvert\cycmat{x}(\sigma)\bigr\rvert\der\sigma.
	\end{equation}
	This completes the proof.\hfill\IEEEQED

\subsection{Spectral correlation of \texorpdfstring{$\{x(n)\}$}{\{x(n)\}} in Section~\ref{sec:numerical}}\label{app:spectral}

	The following is a derivation of the spectral correlation of a \acrshort{pam} signal in \acrshort{wss} noise based on its \acrshort{cl} expansion.
	Refer to~\cite[Ch.~12]{Gardner1988} for an alternative approach.

	The \acrshort{cl} expansion of $\{d(n)\}$ is obtained as follows:
	\begin{equation}
		\begin{aligned}
			& \der\nu_d(f) = \smashoperator[l]{\lim_{N\to\infty}} \frac{1}{N} \mathop{\sum\sum}_{n,k=-\frac{N}{2}}^{\frac{N}{2}-1} a(k) \mathrm{b}(n-kP) \euler^{-\imunit2\pi f n}\\
			&\quad= \Bigl(\sum_{n'}\mathrm{b}(n')\euler^{-\imunit2\pi f n'}\Bigr) \Bigl(\sum_{k}a(k)\euler^{-\imunit2\pi fkP}\Bigr) \der f\\
			&\quad= \mathrm{B}(f) \der f \cdot \sum_{k\in\integs} a(k) \euler^{-\imunit2\pi f kP},
		\end{aligned}
	\end{equation}
	where we have used the change of variable $n'\triangleq n-kP$ and assumed $\sfrac{1}{N}\to\der f$.
	We can now compute the spectral correlation of $\{x(n)\}$ as in~\eqref{eq:spec-corr}:
	\begin{multline}
		\cyc_{X}(\alpha,f) = \cyc_{Z}(\alpha,f) + \mathrm{B}(f) \mathrm{B}^*(f-\alpha) \expec[\euler^{-\imunit2\pi\alpha\varepsilon}] \\
		\cdot \smashoperator{\sum_{k,k'\in\integs}} \expec[a(k)a^*(k')] \euler^{\imunit2\pi(f(k'-k)-\alpha k')P},
	\end{multline}
	due to $\{d(n)\}$ and $\{z(n)\}$ being independent and zero-mean.
	Since symbols are uncorrelated and noise is \acrshort{wss} and white,
	\begin{align}
		&\cyc_{X}(\alpha,f) = \power{z}\delta(\alpha) + \mathrm{B}(f) \mathrm{B}^*(f-\alpha)\frac{1}{\Delta} \int_{\mathrlap{0}}^{\mathrlap{\Delta}} \euler^{-\imunit2\pi\alpha\varepsilon} \der\varepsilon \nonumber\\
		&\quad \cdot \sum_{k,k'} P\delta_{k-k'} \euler^{\imunit2\pi(f(k'-k)-\alpha k')P} \label{eq:spec_corr_sims}\\
		&= \power{z}\delta(\alpha) + \mathrm{B}(f) \mathrm{B}^*(f-\alpha) \frac{1-\euler^{-\imunit2\pi\alpha\Delta}}{\imunit2\pi\alpha\Delta} P \sum_{k\in\integs} \euler^{\imunit2\pi\alpha kP} \nonumber\\
		&= \power{z}\delta(\alpha) + \mathrm{B}(f) \mathrm{B}^*(f-\alpha) \frac{1-\euler^{-\imunit2\pi\alpha\Delta}}{\imunit2\pi\alpha\Delta} \Sha_{\frac{1}{P}}(\alpha).\nonumber
	\end{align}

	\bibliographystyle{IEEEtran}
	\bibliography{IEEEabrv,refs}
	\vspace{-.5em}%
	\begin{IEEEbiography}[{\includegraphics[width=1in,height=1.25in,clip,keepaspectratio]{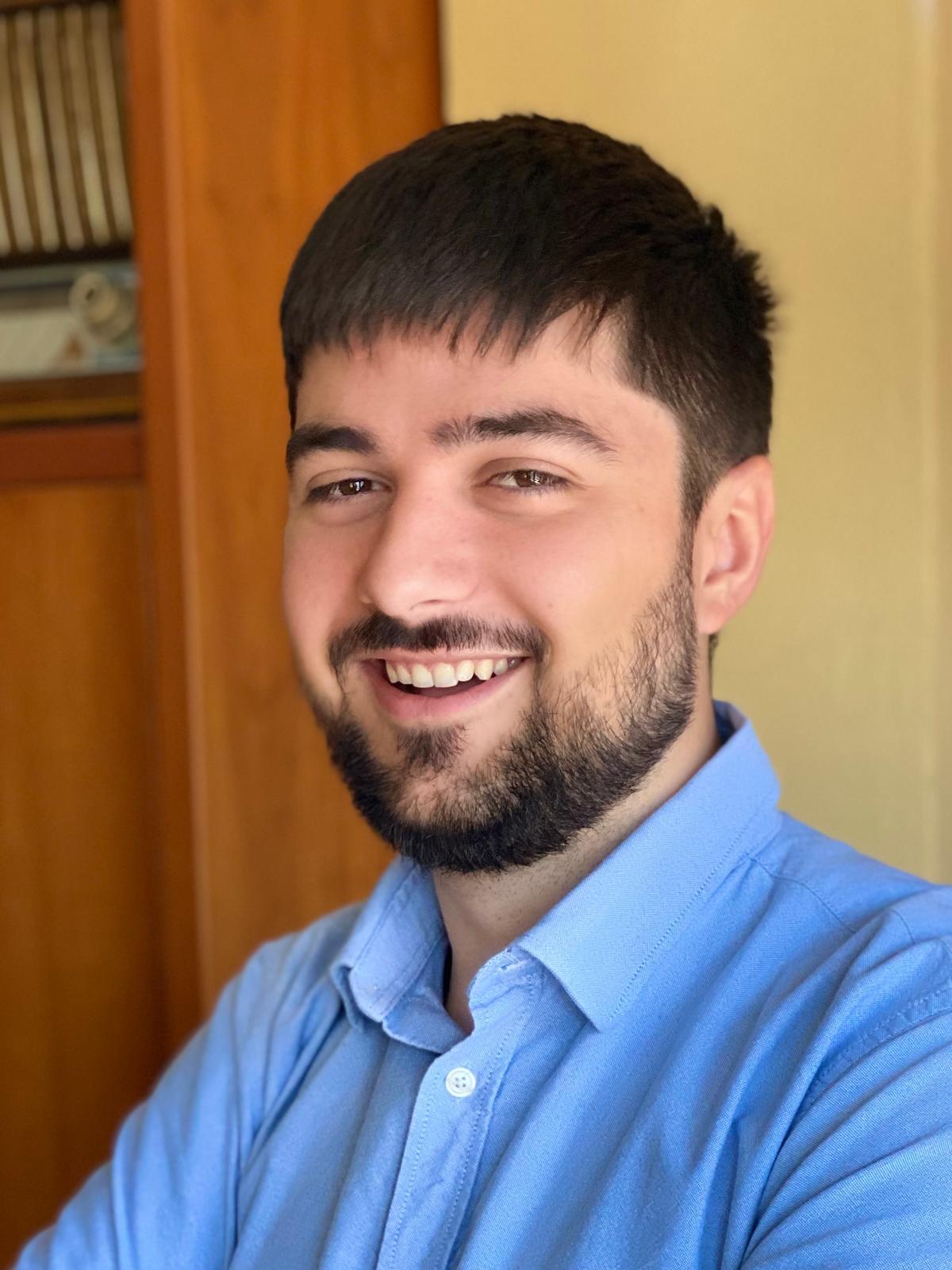}}]{Marc Vilà-Insa}
		(Graduate Student Member, IEEE) received the B.Sc. (2019) and M.Sc. (2021) degrees in Telecommunications Engineering from Universitat Politècnica de Catalunya (UPC), Barcelona.
		He is currently pursuing a Ph.D. degree in Signal Theory and Communications at the UPC, for which he was awarded with predoctoral grant FI-SDUR in 2022, by the Departament de Recerca i Universitats de la Generalitat de Catalunya.
		His areas of expertise are within signal processing for communications.
		His research interests encompass topics related to cyclostationary signal processing, noncoherent detection and massive MIMO communications.
	\end{IEEEbiography}\vspace{-.5em}
	\begin{IEEEbiography}[{\includegraphics[width=1in,height=1.25in,clip,keepaspectratio]{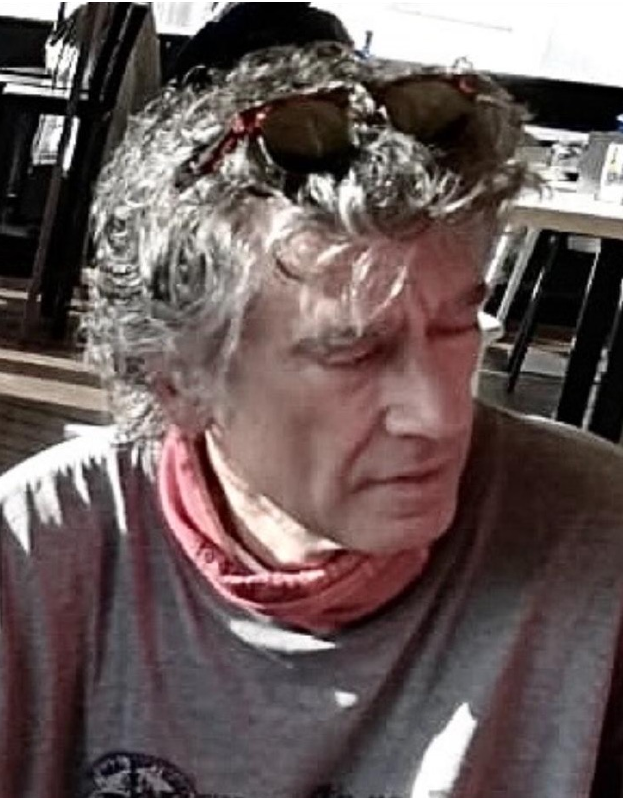}}]{Jaume Riba}
		(Senior Member, IEEE) received the M.Sc and Ph.D. degrees in Telecommunications Engineering from the Universitat Politècnica de Catalunya (UPC), Barcelona.
		He was then promoted in 1997 to Associate Professor in the same alma mater, teaching subjects related to statistical signal processing and digital communication theories.
		He has been regularly involved in research and development programs in the areas of signal processing and satellite communications.
		Along with coauthors in the group, Dr. Riba received the 2003 Best Paper Award of the IEEE Signal Processing Society and 2013 Best Paper Award of the IEEE International Conference on Communications.
		Having research experience in array processing, synchronization, cyclostationarity, source localization, measures of information, sparsity and noncoherent communications, his current research interest is focused on the interplay between information measures and statistical signal processing principles, looking for interpretability in one field in light of the other.
	\end{IEEEbiography}
\end{document}